\documentclass{article}

\usepackage{amsthm}
\usepackage{graphicx}
\usepackage{float}
\usepackage{color}
\usepackage{amsmath}
\usepackage{amssymb}
\usepackage{multirow}
\usepackage{graphicx}
\usepackage{multicol}

\DeclareMathOperator{\sech}{sech}

\begin{document}
\vspace*{0.2in}

\begin{flushleft}
{\Large
\textbf\newline{Interactions of Solitary Pulses of \emph{E. coli} in a One-Dimensional Nutrient Gradient} 
}
\newline
\\
Glenn Young\textsuperscript{1*},
Mahmut Demir\textsuperscript{2}$\dagger$,
Hanna Salman\textsuperscript{2,3},
G. Bard Ermentrout\textsuperscript{4},
Jonathan E. Rubin\textsuperscript{4}
\\
\bigskip
\textbf{1} Department of Mathematics, Eberly College of Science, Pennsylvania State University, State College, PA, USA
\\
\textbf{2} Department of Physics and Astronomy, The Dietrich School of Arts and Sciences, University of Pittsburgh, Pittsburgh, PA, USA
\\
\textbf{3} Department of Computational and Systems Biology, School of Medicine, University of Pittsburgh, Pittsburgh, PA, USA
\\
\textbf{4} Department of Mathematics, The Dietrich School of Arts and Sciences, University of Pittsburgh, Pittsburgh, PA, USA
\bigskip

%
%


$\dagger$ Department of Molecular, Cellular and Developmental Biology, Yale University, New Haven, CT, USA 



* gsy4@psu.edu

\end{flushleft}
\section*{Abstract}
We study an anomalous behavior observed in interacting \emph{E. coli} populations. When two populations of \emph{E. coli} are placed on opposite ends of a long channel with a supply of nutrient between them, they will travel as pulses toward one another up the nutrient gradient.  We present experimental evidence that, counterintuitively, the two pulses will in some cases change direction and begin moving away from each other and the nutrient back toward the end of the channel from which they originated. Simulations of the Keller-Segel chemotaxis model reproduce the experimental results. To gain better insight to the phenomenon,  we introduce a heuristic approximation to the spatial profile of each population in the Keller-Segel model to derive a system of ordinary differential equations approximating the temporal dynamics of its center of mass and  width. This approximate model simplifies analysis of the global dynamics of the bacterial system and allows us to efficiently explore the qualitative behavior changes across variations of parameters, and thereby provides experimentally testable hypotheses about the mechanisms behind the turnaround behavior.

\section*{Author Summary}
The ability of cellular populations to move collectively is necessary for their survival, and consequently studying their collective behavior is important for understanding their life cycle. The mathematical study of such collective behavior has been primarily, though not entirely, focused on the behavior of single population, with little attention given to the interactions of distinct populations of the same species.  We present experimental data demonstrating that, when traveling toward one another in a long channel along a nutrient gradient, two identical populations of \emph{E. coli} will often combine into a single, indistinguishable population, but they can alternatively abruptly stop and change direction without combining, traveling back in the direction from which they respectively came. We study these behaviors by introducing a novel approximation to the spatial profiles of the interacting populations, which allows us to derive a system of ordinary differential equations from a variant of the classical Keller-Segel partial differential equations model. Using this approximation, we analyze how small changes in parameters such as the initial bacterial population sizes, the abundance of nutrient, and the viscosity of the medium in the channel all affect the outcome. Our results will aid the numerical and transient analysis of pulse-pulse interactions in different and more complex environmental conditions.


\section*{Introduction}
Since its inception, the Keller-Segel model for chemotaxis has successfully captured important characteristics of the dynamics of a variety of species, from cellular slime molds such as \emph{Dictyostelium discoideum} to bacteria such as \emph{Escherichia coli} to insects such as the fruit fly \emph{Drosophila melanogaster} \cite{Hillen,Horstmann,Keller1,Keller2,Lof}. Here we use such a model to analyze the transient dynamics of two interacting pulses of bacteria in a one-dimensional nutrient gradient.

 {A bacterial pulse is a common pattern of collective bacterial migration, which has been well characterized in several studies \cite{Ben-Jacob,Emako,Park,Salman2,Saragosti}. Collective behavior and migration of bacteria have been proposed to provide selective advantages \cite{Donlan,Tsimring}. For example,  the increased density associated with congregation supports the formation of biofilms, which provides resistance against antibiotics and other environmental stresses \cite{Davies,Pratt}. Quorum-sensing, a well-known phenomenon in which the gene-expression profile is altered when cells reach critical cell density \cite{Bassler}, is another example where collective migration can aid in reaching the required cell density. Finally, collective migration can be a useful strategy for tracking resources, invading host cells, and responding to an invasion by a competing bacterial population.}

We present experimental results demonstrating the dynamics of two interacting \emph{E. coli} populations in a nutrient gradient. When two \emph{E. coli} populations are placed on opposite ends of a long channel with a supply of nutrient between them, they travel as pulses toward one another up the nutrient gradient.  Interestingly, in some cases they will change direction and begin moving away from each other and the nutrient back towards where they started. Because the two bacterial populations move by chemotaxis up the nutrient gradient and they both produce the same chemoattractant to which they are mutually attracted, it seems reasonable that they should always continue moving inward toward one another, meet in the middle, and subsequently combine into a single, unified population. As this is not the case, we use a Keller-Segel model that includes a  {state variable representing the nutrient supply} to elucidate mechanisms behind this unintuitive direction switch. External gradients have previously been shown to play an important role in collective behavior of species that move by chemotaxis. For example, it has been shown that a  {nutrient} gradient can give rise to a traveling pulse in a population that moves by chemotaxis \cite{Saragosti}. Similar results have been found for temperature and oxygen as well \cite{Demir,Douarche,Salman1,Salman2}.

 {Pulse-pulse interaction has also been studied in a number of reaction-diffusion equations, including the Gierer-Meinhardt and Gray-Scott models \cite{Doelman,Ei,Kang,Sun}. In these works, asymptotic matching is used to derive leading-order ordinary differential equations for the distance between the center of pulses. The stability of the origin of the resulting ODE determines whether the two pulses are predicted to combine or repel.  However, this framework depends heavily on the dynamics of the pulses being slow, and does not allow for analysis of the transient behavior of pulses.} Here we make a straightforward heuristic approximation to the spatial profile of each pulse using a Gaussian distribution. From this assumption, we are able to derive explicit ODEs describing the dynamics of the center of mass and the width of the pulses. In contrast to the Keller-Segel partial differential equation model, our ODE model eases linear stability analysis of equilibrium states and numerical simulation and allows phase plane analysis in some situations. In these ways, our approximate model facilitates analysis of the global dynamics of the bacterial system and enables us to efficiently explore the qualitative behavior changes across variations of parameters.
We show that our approximate model agrees with the Keller-Segel model in predicting that bacterial accumulation is the result of an instability of the uniform state that occurs when the bacterial population size gets sufficiently large. 
Moreover, 
we use our model to analyze parameter conditions that lead to turn around of the bacterial populations and conditions that cause them to combine, obtaining mechanistic predictions for future experimental consideration.

\section*{Results}
We study the interaction of two \emph{E. coli} populations in a one-dimensional nutrient gradient. First, we present experimental results showing two qualitatively different outcomes of bacterial interaction. We then explore the outcomes analytically using both a Keller-Segel model and an ODE model representing pulse dynamics, which we derive.

\subsection*{Experimental Results}\label{sec:exResults}

\begin{figure}[H]
{\centering
\includegraphics[width=3.5in]{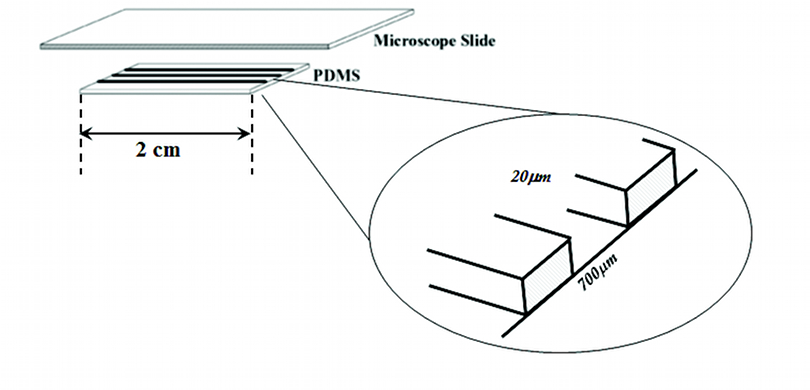}

}
\caption{The experimental setup: a set of narrow channels ($800 \mu$m$\times 20 \mu$m), 2 cm long, microfabricated with polydimethyl-siloxane (PDMS) using the common techniques \cite{Park} and adhered to a microscope slide by plasma cleaning, while leaving both ends open for loading the bacteria.  {After the bacteria are loaded on both ends, the ends of the channels are sealed via epoxy glue.}}
\label{fig:expSetUp}
\end{figure}

\begin{figure}[H]
{ \centering
\includegraphics{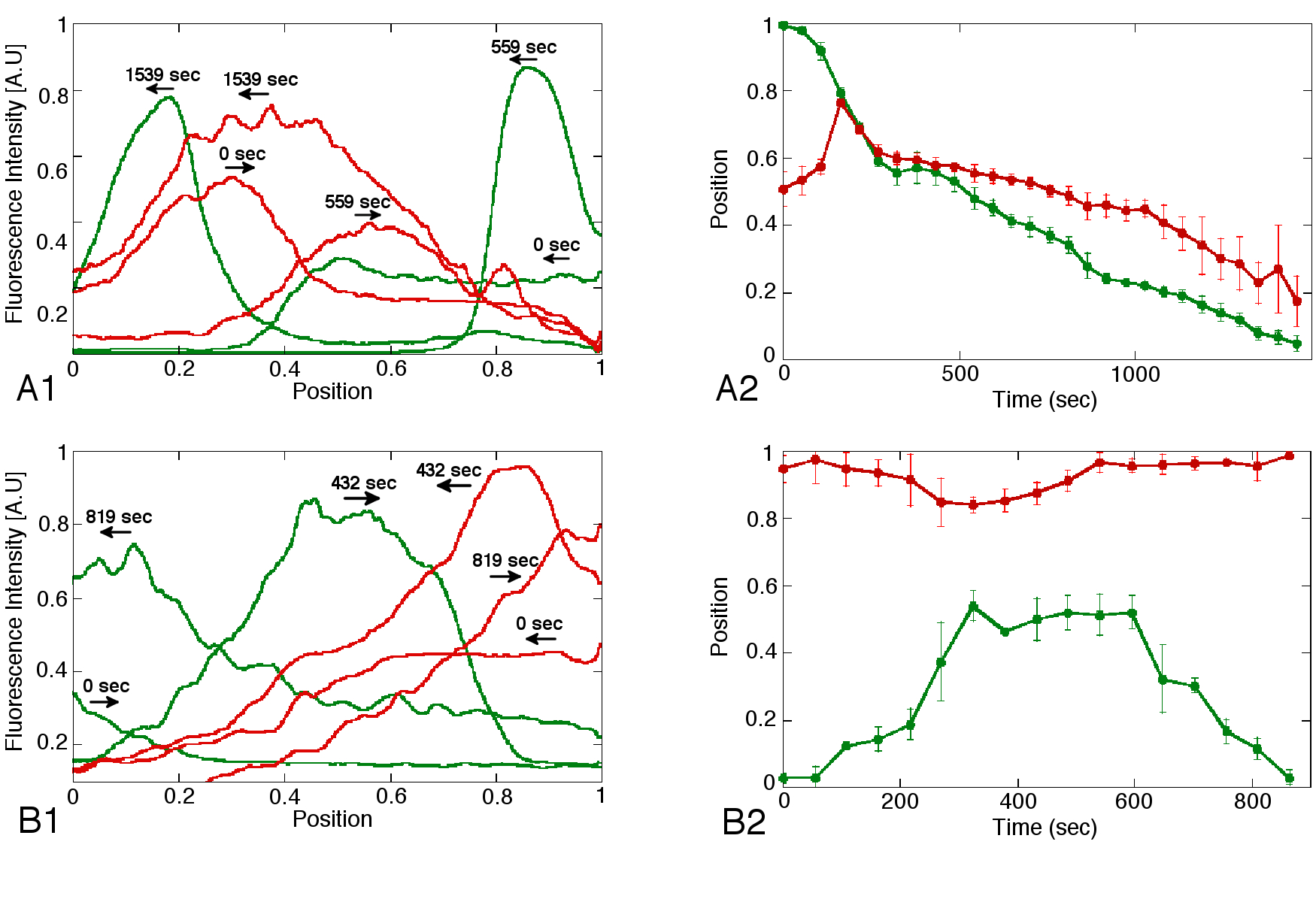}
}
\caption[Experimental results.]{Experimental results.   Examples of the fluorescence intensity profiles along a narrow channel measured for both red and yellow bacteria as indicated by the color of the plot (yellow bacteria are represented by the green line in the figure), at different time points.  {The background fluorescence, which includes some bacteria, was subtracted and the profile was scaled relative to the highest intensity in order to emphasize the pulse itself and to better detect its position.} A1, A2: the two bacterial pulses advance towards each other and when they meet, they combine and move together towards the left end of the channel.  B1, B2: the two pulses bounce back and move towards their original end of the channel. Position  {(horizontal axis in A1, B1, vertical axis in A2, B2) is scaled between 0 and 1} relative to the central 3.5mm of the long axis of the channel  {imaged in the experiment}.  Arrows in A1, B1 indicate direction of movement.  {Note that even though the yellow population in A1 seems to be retreating initially, in reality it is not. The pulse actually arrives from the right, as can be seen clearly by eye during the experiment, but this arrival is not observed in the image due to technical limitations of the imaging process.} Curves in A2, B2 indicate the positions of the centers of mass of the two bacterial populations. The peak position was found by calculating the center of mass of the 10\%  {densest} (with the highest fluorescence intensity) positions for each population. The error bars represent the standard deviation of the center of mass.
  }
  \label{fig:expResults}
\end{figure}

Two bacterial cultures, each expressing a different color fluorescent protein (Red or Yellow), were loaded onto opposite ends of a long narrow channel filled with M9CG medium (Fig \ref{fig:expSetUp}; see Experimental Methods for more details). The bacteria were then observed via fluorescence microscopy at low magnification (2.5x) to allow visualization of collective behavior. Our observations reveal that each bacterial population initially accumulate near the end of the channel forming a sharp concentration peak. The two concentration peaks then proceed to propagate as a pulse towards the center of the channel.  
Subsequently, two possible outcomes were observed (Fig \ref{fig:expResults}A-B).
In the first (Fig \ref{fig:expResults}A), the two populations combine and move together towards one end of the channel  {(observed in {two out of five} experiments)} or sometimes (data not shown) stay at the collision location, while their accumulation peak reduces in amplitude and widens gradually by diffusion  {(observed in {one out of five} experiments)}. In the second case (Fig \ref{fig:expResults}B), the two populations' peaks never meet; rather, they approach each other initially and then bounce back, with each moving towards the end of the channel where it originated  {(observed in {two out of five} experiments)}. {Additional visualization of these behaviors is provided in the Supporting Information.}

\subsection*{ {Agreement of the Keller-Segel model with experimental results}}\label{sec:KSresults}

We use an adaptation of the classic Keller-Segel model of chemotaxis to approximate the spatiotemporal dynamics of the above experiment. We denote the bacterial concentration at time $t\geq0$ and normalized position $0\leq x\leq1$ by $b(t,x)$ and the concentration of chemoattractant by $a(t,x)$. We also include a dynamic variable representing the  {nutrient} concentration, denoted by $\phi(t,x)$. The bacterial concentration diffuses with effective rate $D_b$ and moves by chemotaxis up the chemoattractant gradient at rate $\chi_a$ and up the  {nutrient} gradient with rate $\chi_\phi$.  {In the short observation time of the experiment, the bacterial growth rate is negligible, and we therefore omit it from our model.} The chemoattractant diffuses at rate $D_a$, is produced by the bacteria at rate $r$, and naturally degrades at rate $\delta$. The  {nutrient} has no source, and therefore only diffuses at rate $D_\phi$ and is consumed by the bacteria at rate $\kappa$. Under these assumptions (further detailed in Materials and Methods and Table \ref{tab:params}), the resulting model takes the form

\begin{equation}\label{eq:KS}
\begin{aligned}
\frac{\partial b}{\partial t}&=D_b\frac{\partial^2 b}{\partial x^2}-\chi_a\frac{\partial}{\partial x}\left[b\frac{\partial a}{\partial x} \right]-\chi_\phi\frac{\partial}{\partial x}\left[b\frac{\partial \phi}{\partial x} \right]\\
\frac{\partial a}{\partial t}&=D_a\frac{\partial^2 a}{\partial x^2}+rb-\delta a\\
\frac{\partial \phi}{\partial t} &=D_\phi\frac{\partial^2 \phi}{\partial x^2}-\kappa b\phi,
\end{aligned}
\end{equation}
with no-flux boundary conditions (\ref{eq:BCsND}).

 {Because we seek to capture the experimentally observed dynamics of the bacterial pulses using a Keller-Segel model, we split the bacterial population $b$ into two sub-populations, $b$ and $\beta$, both of which produce chemoattractant $a$ and $
\alpha$, respectively, to which they are mutually attracted. The model we consider is therefore
\begin{equation}\label{eq:KS2}
\begin{aligned}
\frac{\partial b}{\partial t}&=D_b\frac{\partial^2 b}{\partial x^2}-\chi_a\frac{\partial}{\partial x}\left[ b\frac{\partial (a_1+a_2)}{\partial x}\right]-\chi_\phi\frac{\partial}{\partial x}\left[ b\frac{\partial \phi}{\partial x}\right]\\
\frac{\partial \beta}{\partial t}&=D_b\frac{\partial^2 \beta}{\partial x^2}-\chi_a\frac{\partial}{\partial x}\left[ \beta\frac{\partial (a_1+a_2)}{\partial x}\right]-\chi_\phi\frac{\partial}{\partial x}\left[ \beta\frac{\partial \phi}{\partial x}\right]\\
\frac{\partial a_1}{\partial t}&=D_a\frac{\partial^2 a_1}{\partial x^2}+rb-\delta a_1\\
\frac{\partial a_2}{\partial t}&=D_a\frac{\partial^2 a_2}{\partial x^2}+r\beta-\delta a_2\\
\frac{\partial \phi}{\partial t} &=D_\phi\frac{\partial^2 \phi}{\partial x^2}-\kappa (b+\beta)\phi
\end{aligned}
\end{equation}
with boundary conditions (\ref{eq:BCs2ND}).}
The dynamics of this two-population model are identical to the dynamics of the single-population model because of the linearity of system (\ref{eq:KS}) in $b$. Explicitly separating the two populations allows us to easily distinguish between bacteria originating in different pulses, however. We initialize all simulations with the two populations accumulated on opposite ends of the spatial domain. We assume that sufficient time has passed so that the bacteria have consumed  the nutrient  at the densely populated regions at the ends of the domain so that the nutrient concentration is initially distributed as the symmetric sigmoid function given by (\ref{eq:initFood}) in Materials and Methods. Without the  {nutrient}, the bacterial populations would remain accumulated at their respective ends of the domain, maintaining a concentration of chemoattractant, and would not travel inward.

 {In order to form and maintain a pulse, the bacterial population must exceed a critical threshold \cite{Salman2}. If the cell density is too low, the secreted attractant cannot accumulate fast enough to achieve the minimal concentration required for other bacteria to sense it. The Keller-Segel model captures this phenomenon: cellular populations evolving according to this model can only form a nontrivial pulse if the population size is sufficiently large relative to model parameters \cite{Keshet, Hillen, Horstmann,Murray}.} Below this critical threshold, the only solution is the uniform solution, $b=b_{tot}= constant$, $a=rb_{tot}/\delta$. Model (\ref{eq:KS}) in particular predicts that in order to maintain a nontrivial pulse, the total amount of bacteria must be greater than the critical threshold defined by
\begin{equation}\label{eq:KSthresh}
b_{tot}^*=\frac{D_b[\pi^2D_a+\delta]}{r \chi_a}
\end{equation} \cite{Keshet}. 
The nutrient does not  {affect the asymptotic stability of the uniform state because it vanishes at steady state. The presence of nutrient can cause transient bacterial pulses to form, but the system will always settle into the uniform state when the total amount of bacteria is below the threshold $b_{tot}^*$ specified in equation (\ref{eq:KSthresh}).}

\begin{figure}[H]
{ \centering
\includegraphics[width=3in]{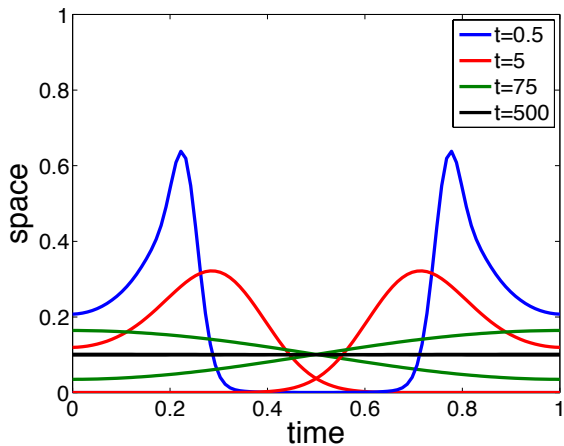}

}
\caption[Bacteria dynamics predicted by the Keller-Segel model with population size below critical threshold (\ref{eq:KSthresh}).]{Bacteria dynamics predicted by the Keller-Segel model with population size below critical threshold (\ref{eq:KSthresh}). The two populations originating at opposite domain boundaries initially move up the nutrient gradient but cannot maintain pulse-like profiles.}
\label{fig:PDEdiffuse}
\end{figure}

Below threshold (\ref{eq:KSthresh}), the bacterial population cannot maintain a pulse-like solution. Fig \ref{fig:PDEdiffuse} shows an example of a simulation of model (\ref{eq:KS2}) when the combined bacterial population size is less than threshold (\ref{eq:KSthresh}).  The two populations initially form pulses and move up the food gradient toward the center, but eventually lose their pulse-like shapes and diffuse out to uniformly fill the spatial domain.  {This is consistent with the experimentally observed (but not shown here) outcome in which the two bacterial populations combine in the interior of the domain, remain stationary, and slowly approach a uniform distribution through diffusion.}

When the total amount of bacteria exceeds threshold (\ref{eq:KSthresh}), the bacterial populations will asymptotically form a pulse along one or both of the boundaries of the spatial domain.  {If both populations fall below the threshold and are sufficiently far apart in space, neither will be able to maintain a pulse-like distribution. However, even if the two populations are individually below threshold (\ref{eq:KSthresh}), they can still form a pulse if they are close in space and their combined population exceeds this value.} For consistency with experiments, we will henceforth only consider bacterial population sizes above this threshold. 

\begin{figure}[H]
\centering \includegraphics{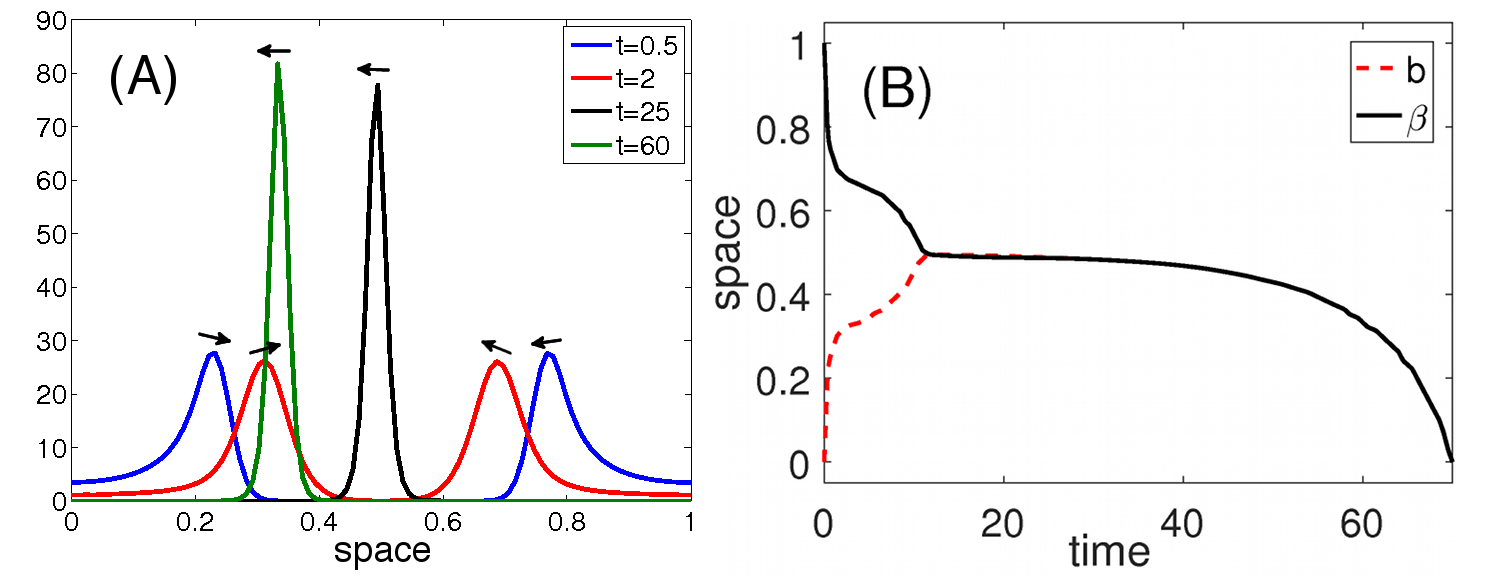}
\caption[Bacterial pulses combining under the dynamics of system (\ref{eq:KS2}).]{Bacterial pulses combining under the dynamics of system (\ref{eq:KS2}) with boundary conditions (\ref{eq:BCs2ND}).  
The initial food profile is given by the reflected sigmoid (\ref{eq:initFood}) with $\phi_0=20$. (A) Snapshots of the bacterial profiles at different times. The arrows indicate direction of motion. By $t=25$, the two populations have combined and begun moving toward the left boundary. (B) The positions of the peaks of the bacterial pulses over time.}
\label{fig:PDECombine}
\end{figure}

\begin{figure}[H]
\centering \includegraphics{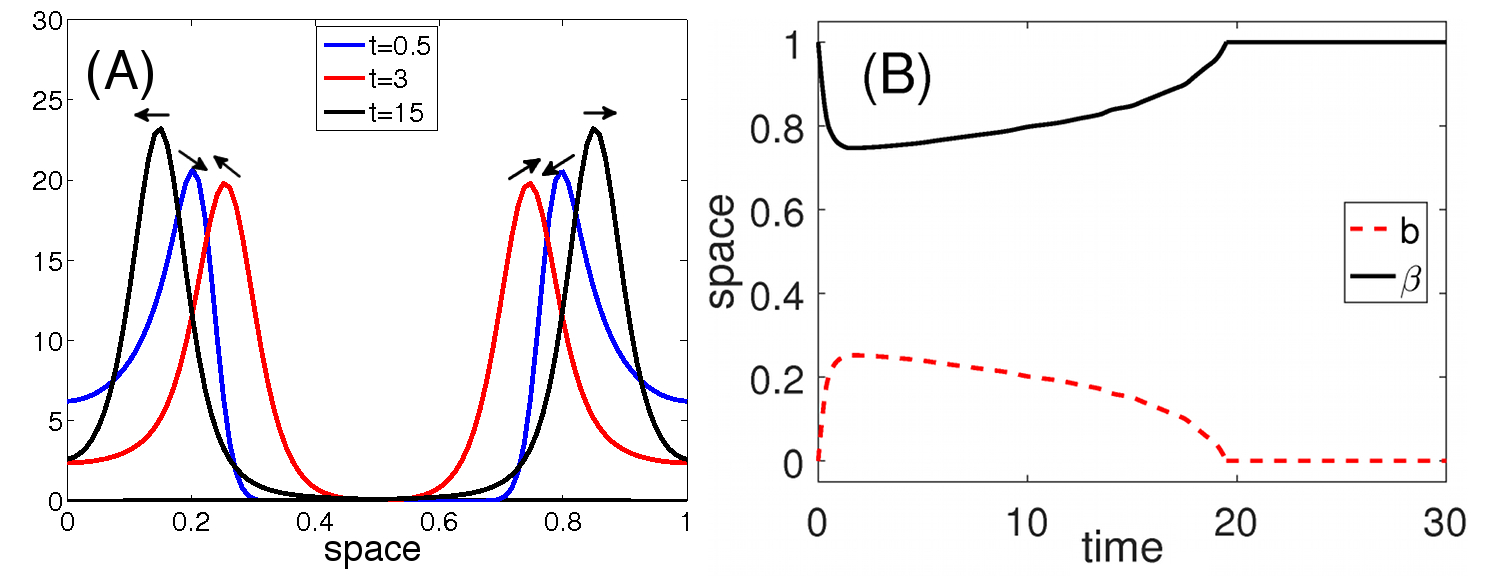}
\caption[Bacterial pulses turning around under the dynamics of system (\ref{eq:KS2}).]{Bacterial pulses turning around under the dynamics of system (\ref{eq:KS2}) with boundary conditions (\ref{eq:BCs2ND}). 
The initial food profile is given by the reflected sigmoid (\ref{eq:initFood}) with $\phi_0=18$. (A) Snapshots of the bacterial profiles at different times. Arrows indicate direction of motion. (B) The positions of the peaks of the bacterial pulses over time.}
\label{fig:PDETurn}
\end{figure}

Figs \ref{fig:PDECombine} and \ref{fig:PDETurn} show examples of simulations of system (\ref{eq:KS2}) that capture the two qualitatively distinct results observed experimentally. In Fig \ref{fig:PDECombine}, the two bacterial populations move up the nutrient gradient toward one another until they meet and combine into a single pulse, which propagates to  {the left} end of the domain.  {The direction of propagation after combination here is due to a small asymmetry in the initial conditions: the population beginning on the left is slightly smaller than the population beginning on the right. Without this asymmetry, the two populations would simply remain as a pulse in the center of the domain after combination.} In Fig \ref{fig:PDETurn}, the two populations initially move up the  {nutrient} gradient but eventually change direction and move backwards toward the chemoattractant that is accumulated near the boundaries. The only difference between the two outcomes is the initial amount of nutrient: the simulations shown in Fig \ref{fig:PDECombine} begin with more nutrient than those  in Fig \ref{fig:PDETurn} {(see Supporting Information for visualization of the chemoattractant and nutrient profiles).}

Though the distinction between these outcomes results from a change in the initial abundance of the nutrient, we note that we can produce similar results by changing other model parameters. For example, if we start from conditions that result in the two populations combining, reducing the total size of both bacterial populations results in both populations turning around (data not shown). The population size can feasibly vary between experiments, and is therefore another potential factor in determining whether the populations will combine or turn around.

We seek to determine possible causes of these distinct outcomes. We observe that both the combining outcome and the turnaround outcome can be characterized by the relative position of the center of mass of the two bacterial populations: if the centers of mass coalesce, then the two populations have combined; if they change direction and accumulate along the opposite boundaries of the domain, then the two populations have turned around. 
We next derive a system of ordinary differential equations (ODEs) describing the dynamics of the size, center of mass, and variance of the spatial profile of each bacterial pulse. 
By maintaining information about these critical characteristics with an ODE setting, we can much more efficiently explore and generate predictions about parameter-dependence of solution behavior.

\subsection*{Gaussian approximation to the spatial distribution}\label{sec:gaussResults}

Here we consider the temporal dynamics of the spatial moments of each variable in a  Keller-Segel model for one bacterial population (\ref{eq:KS}). The $i^{th}$ moment, $i=0,1,2,\dots$, of the spatial profile of variable $s\in\{b,a,\phi\}$ is defined by 
\begin{equation*}
s_i(t)=\displaystyle \int_{-\infty}^{\infty} x^is(x,t)dx.
\end{equation*}
Each variable represents the profile of a particular model component, and we are primarily concerned with the total availability, center of mass, and variance or spread of each variable. The  zeroth moment of variable $s$, or $s_0$, quantifies the amount of the corresponding component that is present. The center of mass for each variable is given by the formula $\mu_s=s_1/s_0$, and the variance is given by $\sigma_s^2=s_2/s_0-\mu_s^2$. Each of these quantities is a function of time only, and therefore differentiation produces an ordinary differential equation describing the temporal dynamics of that quantity. The resulting system is given by (\ref{eq:gauss}) in Materials and Methods. System (\ref{eq:gauss}) depends on several mixed moments; for example, the differential equation for $\mu_b$ is

\begin{equation}\label{eq:momentEx}
\dot\mu_b=\frac{\chi_a\langle ba_x\rangle+\chi_\phi\langle b\phi_x\rangle}{b_0},
\end{equation}
where $$\langle f(x)g(x)\rangle=\displaystyle \int_{-\infty}^\infty f(x)g(x)dx.$$
 The two terms on the right hand side of equation (\ref{eq:momentEx}), $\chi_a\langle ba_x\rangle$ and $\chi_\phi\langle b\phi_x\rangle$, are mixed moments, which cannot be found without knowing the spatial distribution of each variable. Indeed, the differential equation for each moment of each variable will generally depend on higher or mixed moments, and we therefore require a method of moment closure.

We observe that the spatial profiles of the bacterial populations in the Keller-Segel model (\ref{eq:KS2}) both maintain pulsatile, fairly symmetric, Gaussian-like appearances when accumulated in the interior of the spatial domain (see Figs \ref{fig:PDECombine} and \ref{fig:PDETurn}). We therefore approximate the spatial distribution of each variable $s(t,x)\in\{b(t,x),a(t,x),\phi(t,x)\}$ by
\begin{equation}\label{eq:gaussProfile}
s(t,x)=\frac{s_0}{\sigma_s\sqrt{\pi}}\exp\left(\frac{-(x-\mu_s)^2}{\sigma_s^2}\right).
\end{equation}
Further, because the pulse-pulse interaction occurs within the interior of the domain, we ignore boundary effects by considering the system on the infinite real line. Approximation (\ref{eq:gaussProfile}) allows us to evaluate each integral that appears in system (\ref{eq:gauss}), resulting in an explicit system of ordinary differential equations. In this way, the approximation allows us to close our system of moment equations.


The behaviors of the dynamic variables in system (\ref{eq:gauss}) describe important aspects of the dynamics of the populations considered in the Keller-Segel model (\ref{eq:KS}). In particular, a change in the direction of $\mu_b$ (that is, the sign of $\dot{\mu}_b$) corresponds to a direction reversal of the bacterial population. Similarly, if $\sigma_b^2$ is nonzero and small, then the bacteria form a pulse; if $\sigma_b^2$ tends to infinity, then the bacterial population diffuses out to a uniform state.

Before we explore the mechanisms responsible for the turnaround of the bacteria in a one-dimensional nutrient gradient, we explore the extent of the qualitative agreement between the Keller-Segel model (\ref{eq:KS}) and model (\ref{eq:gauss}).


\subsubsection*{Comparison with Keller-Segel model}\label{sec:gaussCompare}
A foundational result for the original Keller-Segel model is its ability to explain bacterial pulse formation as a Turing instability of the uniform state. In particular, the size of the bacterial population must exceed threshold (\ref{eq:KSthresh}) in order for the bacteria to form and maintain a pulse. Model (\ref{eq:gauss}) predicts a similar threshold that the bacterial population size must overcome in order to form a nontrivial pulse (see Materials and Methods): $$b_0^*=\frac{D_b\sqrt{54\pi D_a\delta}}{r\chi_a }.$$

\begin{figure}[H]
{\centering
\includegraphics{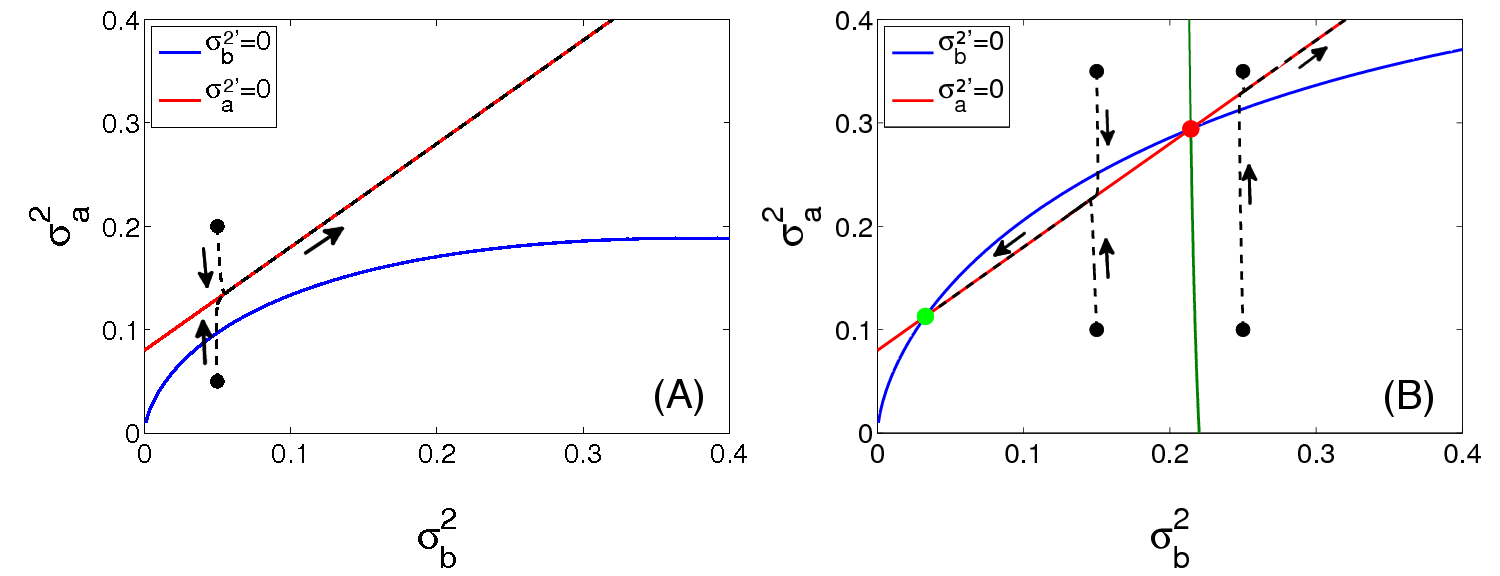}

}
\caption[Phase plane of system (\ref{eq:ABvar}) describing steady states of the Gaussian approximation model.]{Phase plane of system (\ref{eq:ABvar}) describing steady states of the Gaussian approximation model.  (A) The bacterial size $b_0=0.008$ is below the critical threshold $b_0^*$. Trajectories approach the $\sigma_a^2$-nullcline (red) and then both $\sigma_a^2$ and $\sigma_b^2$ tend to infinity. (B) The bacterial population size $b_0=0.012$ is above the critical threshold $b_0^*$. The left-most equilibrium point is a stable node (green dot). The right-most equilibrium point is a saddle (red dot), the stable manifold of which is shown as the green curve. To the left of this stable manifold, trajectories tend toward the stable node, and the bacteria consequently form a pulse. To the right of the manifold, trajectories tend to infinity, and the bacteria diffuse out to the uniform solution. Arrows in both panels indicate the direction of flow and the black dots indicate representative initial conditions.}
\label{fig:nullclines1}
\end{figure}

\begin{figure}[H]
{\centering
\includegraphics[width=2.5in]{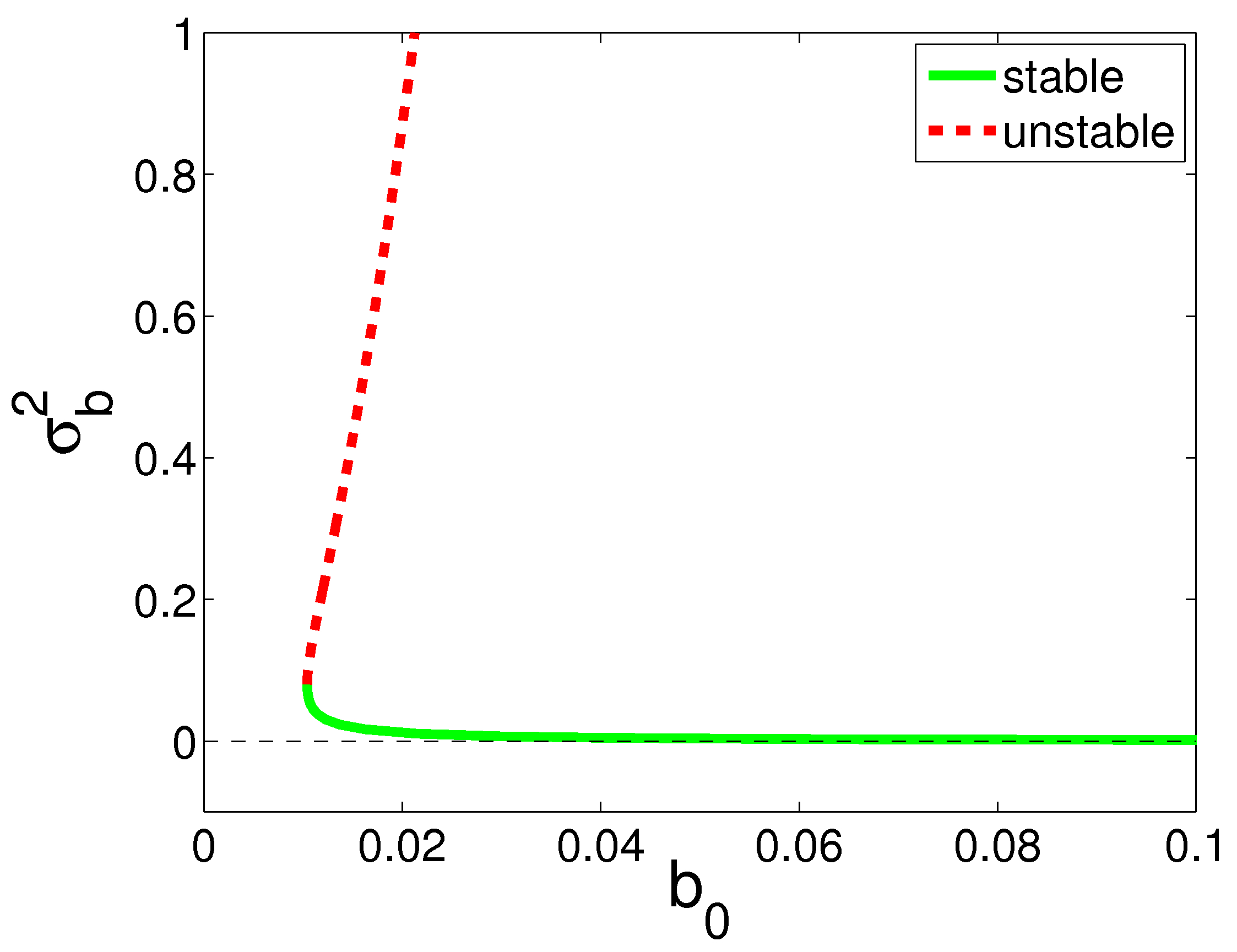}

}
\caption[Bifurcation diagram of system (\ref{eq:ABvar}).]{Bifurcation diagram of system (\ref{eq:ABvar}). The green curve corresponds to the $\sigma_b^2$ coordinate of the stable node, and the red dashed curve corresponds to the $\sigma_b^2$ coordinate of the saddle point. The $\sigma_b^2$ coordinate of the saddle point increases with $b_0$, and consequently the separatrix in Fig \ref{fig:nullclines1} gets pushed farther to the right, promoting pulse formation.}
\label{fig:varBif1}
\end{figure}

Fig \ref{fig:nullclines1} shows the $\sigma_b^2$, $\sigma_a^2$ phase plane when $b_0$ is below (A) and above (B) threshold $b_0^*$.  When $b_0<b_0^*$, both $\sigma_b^2$ and $\sigma_a^2$ tend to infinity for all initial conditions; that is, the bacterial population cannot form a pulse and will asymptotically diffuse to the uniform state. As $b_0$ becomes larger than $b_0^*$, the system undergoes a saddle-node bifurcation, at a value computed analytically in Materials and Methods, and the bacteria are able to form and maintain a nontrivial pulse-like solution.  However, the asymptotic states are now bistable. If $\sigma_b^2$ is initially too high (to the right of the green curve in Fig \ref{fig:nullclines1}B), the bacteria will simply diffuse out to a uniform state. In this way, our model (\ref{eq:gauss}) differs from the Keller-Segel model (\ref{eq:KS}): sufficiently small perturbations off of the uniform state will not result in bacterial pulse formation. However, as $b_0$ increases, the $\sigma_b^2$ coordinate of the saddle point rapidly increases (Fig \ref{fig:varBif1}). Consequently, the initial variance $\sigma_b^2$ required for the bacteria to asymptotically remain uniform can be made arbitrarily large by making $b_0$ sufficiently large. 
 Furthermore, the threshold $b_0^*$ agrees qualitatively with the analogous threshold given in equation (\ref{eq:KSthresh}) for system (\ref{eq:KS}): increasing $D_b$, $D_a$, or $\delta$ increases the threshold, while increasing $r$ or $\chi_a$ decreases it. Thus, the two models generally yield qualitatively similar predictions about the effects of parameters on the requirements for pulse formation.

\subsection*{Predicting turnaround}

As with model (\ref{eq:KS}), we will only consider regimes under which the bacterial population maintains a size above the critical threshold $b_0^*$ and can therefore maintain a pulse. We now consider a two-population version of the Gaussian approximation model, given by system (\ref{eq:gauss2}) (see Materials and Methods). This model is derived analogously to model (\ref{eq:gauss}) but includes two separate bacterial populations $b$ and $\beta$ with separate chemoattractant densities $a$ and $\alpha$, respectively. The dynamics governing $b$ and $\beta$ are identical: both are mutually attracted by chemotaxis up both chemoattractant gradients and the nutrient gradient and they diffuse at the same rate.

\begin{figure}[H]
\centering \includegraphics{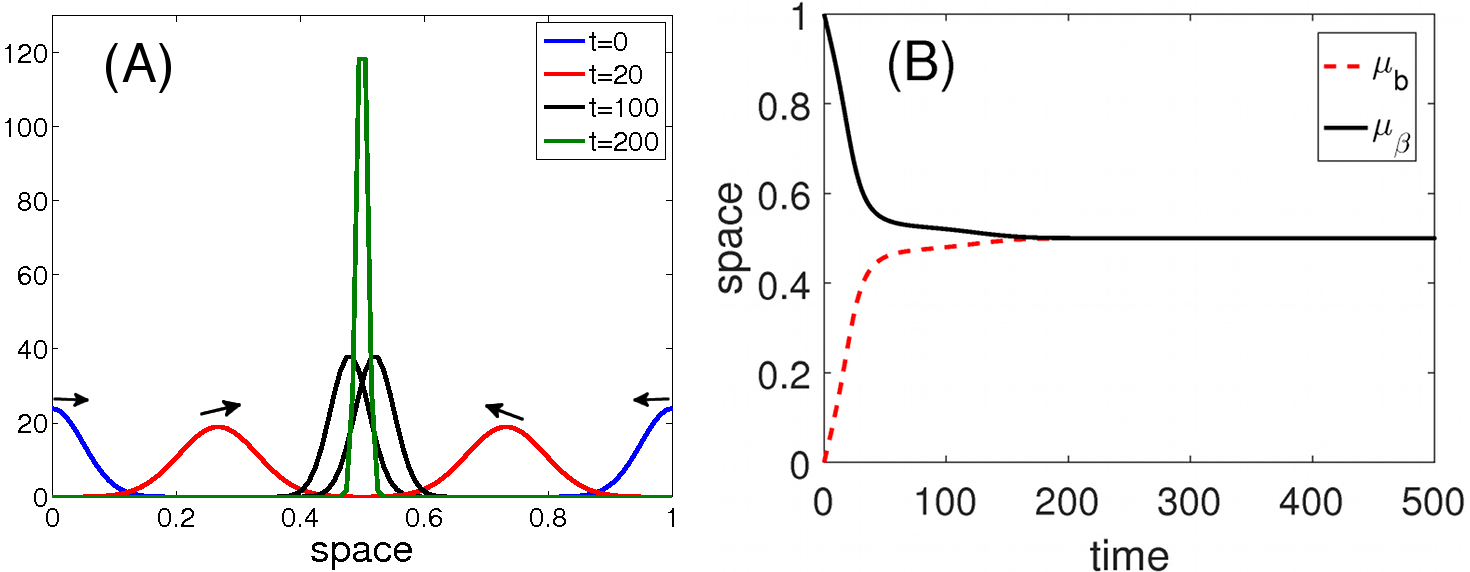}
\caption[Bacterial pulses combine under the dynamics of system (\ref{eq:gauss2}).]{Bacterial pulses combine under the dynamics of system (\ref{eq:gauss2}). The two populations move toward one another up the  {nutrient} gradient until they collide and combine to form a single pulse. The initial amount of nutrient is $\phi_0(0)=35$. (A) Snapshots of the bacterial spatial profiles at different times, given by (\ref{eq:gaussProfile}). The arrows indicate direction of motion. (B) The positions of the peaks of the bacterial pulses over time.}
\label{fig:momentsCombine}
\end{figure}

\begin{figure}[H]
\centering \includegraphics{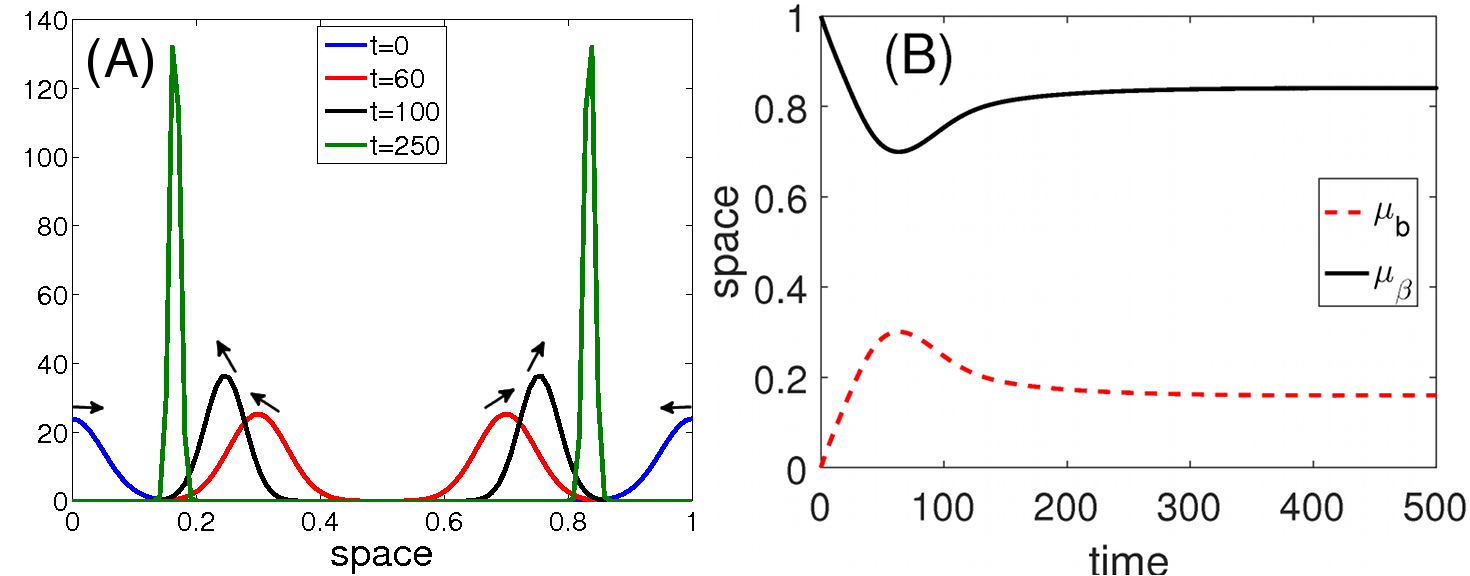}
\caption[Bacterial pulses turn around under the dynamics of system (\ref{eq:gauss2}).]{Bacterial pulses turn around under the dynamics of system (\ref{eq:gauss2}). The two populations initially move toward one another up the  {nutrient} gradient but later change direction and move back toward their own accumulated chemoattractant. The initial amount of nutrient is $\phi_0(0)=25$. (A) Snapshots of the bacterial spatial profiles (\ref{eq:gaussProfile}) at different times. The arrows indicate direction of motion. (B) The positions of the peaks of the bacterial pulses over time.}
\label{fig:momentsTurn}
\end{figure}

 Two example outcomes of simulations of model (\ref{eq:gauss2}) demonstrating agreement with the experimentally observed outcomes are shown in Figs \ref{fig:momentsCombine}  and \ref{fig:momentsTurn}. The only difference between the two simulations was the initial condition for shared nutrient, $\phi_0$. In Fig \ref{fig:momentsCombine}, $\phi_0(0)=35$,  {and the two populations combine and remain stationary after combination. This is in contrast with Fig \ref{fig:PDECombine}, in which the combined population moved toward the left end of the domain. System (\ref{eq:gauss2}) is on the spatial domain $(-\infty,\infty)$, however, and the combined population therefore does not have a boundary to travel toward, and remains stationary once the centers of mass of the two bacterial populations and their respective chemoattractant densities coalesce (see Section Capturing turnaround in Materials and Methods).} In Fig \ref{fig:momentsTurn}, $\phi_0(0)=25$, and the two populations turn around.   The parameters chosen in both simulations are those in Table \ref{tab:GaussParams}. The results shown in Figs \ref{fig:momentsCombine} and \ref{fig:momentsTurn} are consistent with the results from simulations of the Keller-Segel model (\ref{eq:KS2}): increasing the initial amount of  {nutrient} causes the bacteria to switch from a regime in which they turn around to one in which they combine (Figs \ref{fig:PDECombine} and \ref{fig:PDETurn}).  {We note that while we have assumed that the two population profiles are initially symmetric in size and width, small asymmetries do not affect the qualitative transient behavior (results not shown).}

Our simulations confirm that a change to the initial amount of available nutrient can cause a change in the outcome of the bacterial interaction. Variations in other parameters and initial conditions, such as the diffusivity of the medium in the channel and the initial size of the bacterial populations, can have similar effects on the behavior of the bacterial pulses. In considering whether a bacterial population turns around, we are particularly concerned with the behavior of the center of mass of the population, $\mu_b$. One significant advantage of model (\ref{eq:gauss2}) over the two-population Keller-Segel model (\ref{eq:KS2}) in this regard is that model (\ref{eq:gauss2}) explicitly includes the time derivative of this center of mass and thus allows us to separately consider the effects that the chemotactic attraction to the chemoattractant and to the  {nutrient} have on its motion.


The differential equation for $\mu_b$ is

{
\begin{equation}\label{eq:mub}
\begin{aligned}
\dot{\mu}_b &=\frac{\chi_a\langle ba_x\rangle+\chi_a\langle b\alpha_x\rangle+\chi_\phi\langle b\phi_x\rangle}{b_0}\\
&=  -\frac{2\chi_aa_0(\mu_b-\mu_a)}{\sqrt{\pi}(\sigma_b^2+\sigma_a^2)^{3/2}}\exp\left(\frac{-(\mu_b-\mu_a)^2}{\sigma_b^2+\sigma_a^2}\right)\\
 &{\color{white}=} - \frac{2\chi_a\alpha_0(\mu_b-\mu_\alpha)}{\sqrt{\pi}(\sigma_b^2+\sigma_\alpha^2)^{3/2}}\exp\left(\frac{-(\mu_b-\mu_\alpha)^2}{\sigma_b^2+\sigma_\alpha^2}\right)\\
 &{\color{white}=}  -\frac{2\chi_\phi\phi_0(\mu_b-\mu_\phi)}{\sqrt{\pi}(\sigma_b^2+\sigma_\phi^2)^{3/2}}\exp\left(\frac{-(\mu_b-\mu_\phi)^2}{\sigma_b^2+\sigma_\phi^2}\right),
\end{aligned}
\end{equation}
}where each of the three terms in the sum in the right hand side of (\ref{eq:mub}) can be interpreted, in order, as the rate of change in position of the center of mass of $b$ due to its own chemoattractant, due to the other population's chemoattractant, and due to the  {nutrient}, respectively. If we denote the distance between the center of mass of the bacterial pulse $b$ and that of any of the attracting substances by $x$ and the sum of the variances of the bacterial population and the same substance by $y$, then the rate of change of $\mu_b$ due to that substance is of the form $x/y^{3/2}\exp(-x^2/y).$  Thus, if the centers of mass of the bacterial pulse and an attractant lie at the same position, then  the bacterial population experiences  no chemotactic pull due to that attractant. If the distance between the pulses is large, the chemotactic pull is exponentially small but nonzero. The chemotactic attraction is maximized with respect to distance $x$ when the two pulses are a small but nonzero distance away from one another. For small variances, the two pulses must be very close in order to provide a large chemotactic attraction. This makes sense, since if both populations are accumulated in very tight pulses, the overlap between the two will be minimal, and consequently the chemical gradient sensed by the bacteria will be small. If the variance of either population is large, however, then even if the distance between the two pulses is large, the pulses can overlap nontrivially, and the bacteria will be attracted up the chemical gradient.

Since we assume that $\mu_b(0)=\mu_a(0)=0$, $\mu_\phi(0)=0.5$, and $\mu_\beta(0)=\mu_\alpha(0)=1$, we have that $\dot{\mu}_b(0)>0$ and $\dot{\mu}_a(0)=0$. The center of mass of the bacteria is therefore generically ahead (with respect to the direction of motion) of the center of mass of the chemoattractant for early time. We can now apprehend the mechanism that allows for the bacteria to turn around: the bacteria are attracted inward toward the nutrient and the second population's chemoattractant and outward by their own chemoattractant. If the outward attraction becomes stronger than the inward attraction, then the bacteria will turn around.

Upon inspecting equation (\ref{eq:mub}), it is clear that the chemotactic pull toward any given substance is related to the chemotactic sensitivity to the substance ($\chi_a$ and $\chi_\phi$), the distance between the center of mass of the bacterial population and that of the substance, the variance of the pulse of the substance, and the total amount of the substance present. Since the latter three quantities are dynamic variables, direct analysis of their effects on the transient behavior of $\mu_b$ is not viable. Instead, we consider the effects of parameters related to the dynamics of these variables.

\begin{figure}[H]
{\centering
 \includegraphics{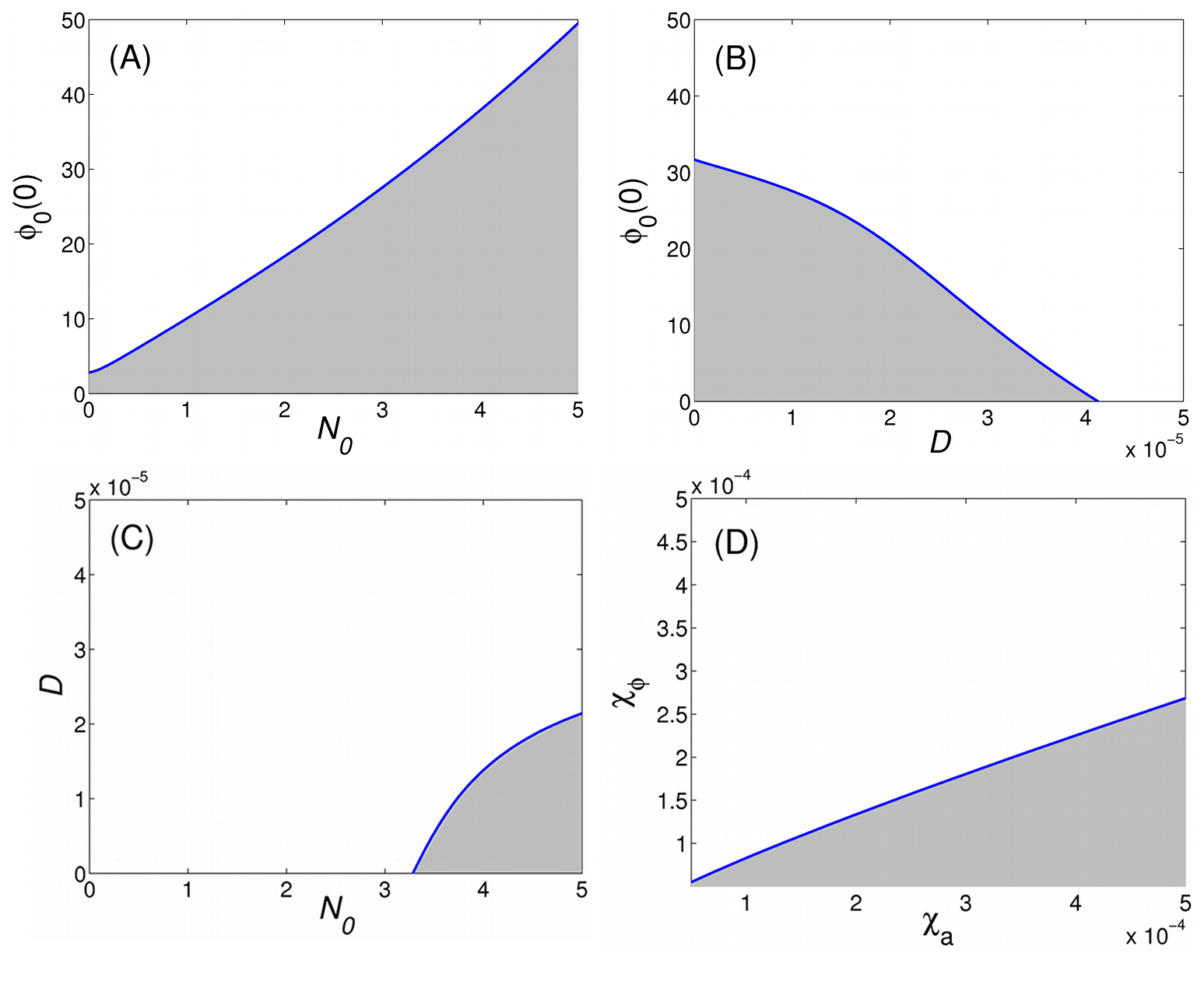}

}
\caption[Boundaries in parameter space between combination (white) and turnaround (grey).]{Boundaries in parameter space between combination (white) and turnaround (grey). The criterion for combination is $\mu_b-\mu_\beta=0$ before $t=500$. Whenever a parameter is not varied, the common bacterial population size $N_0=3$, while $D=10^{-5}$, $\chi_a=0.00025$, $\chi_\phi=0.0002$, and $\phi_0(0)=35$.}
\label{fig:combine}
\end{figure}

We harness the tractable ODE model to develop a boundary value problem that determines the boundary between regimes where bacteria reverse direction and those where bacteria combine in various parameter spaces. Fig \ref{fig:combine} shows the results of solving this boundary value problem.  In this figure, we denote the equal size of the two bacterial populations by $N_0:=b_0=\beta_0$, the diffusivity of the bacteria by $D=D_b$, and we assume that $D_a=D_\phi=20\times D$. Each panel shows a given parameter space divided into two regions. Parameter pairs chosen from the grey region in each panel represent a regime in which the two \emph{E. coli} populations turn around; parameters chosen from the white region correspond to a regime in which they combine. These figures provide a picture of the relative contributions of  the parameters considered. For example, Fig \ref{fig:combine}A shows that if the bacterial population size is increased, more  {nutrient} is needed to result in the bacterial populations combining. This is easy to understand: if the bacterial populations are larger, then they produce more chemoattractant, and the outward attraction toward the bacteria's own chemoattractant will be stronger, such that combination requires a stronger inward attraction toward the nutrient.

Figs \ref{fig:combine}B and C are more subtle. Increasing $D$ can be interpreted as, for example, decreasing the viscosity of the medium in which the bacteria are suspended, thereby increasing the diffusivity of the bacterial and chemical populations. Fig \ref{fig:combine}B shows that the higher the diffusion rate, the less initial nutrient is necessary to cause the bacterial population to combine. For too fluid of a medium, the chemoattractant of both populations spreads quickly across the spatial domain to reach the other population.  This results in a mutual attraction of both populations toward one another, and the  {nutrient} is no longer needed to pull both populations inward. Fig \ref{fig:combine}C similarly shows that in order for the two populations to turn around when diffusivity is high, they need a large initial population  size, which provides a large initial supply of chemoattractant.

Fig \ref{fig:combine}D shows the chemotactic sensitivity of the bacteria toward the chemoattractant, $\chi_a$, versus the sensitivity toward the nutrient, $\chi_\phi$. While these parameters do not change across experimental trials, this figure is easily interpreted and agrees with intuition: a strong attraction toward the nutrient will always result in the bacterial populations being pulled quickly inward and combining. If the attraction toward the chemoattractant is sufficiently high relative to the attraction toward the nutrient, then the bacteria will be pulled strongly outward toward the previously accumulated chemoattractant  located closer to the domain boundaries and hence will turn around.


\section*{Discussion}
In this paper, we study the interaction of bacterial pulses in a one-dimensional nutrient gradient. We present experimental results in which two identical populations of \emph{E. coli} moving toward one another up a nutrient gradient change direction and move back in the directions from which they came, rather than continuing toward each other to combine into one unified population. We capture this turn-around behavior numerically using the classic Keller-Segel model for bacterial chemotaxis. We then develop a system of ordinary differential equations approximating the spatiotemporal dynamics of the spatial moments of each bacterial population and associated attractants represented in the Keller-Segel model. Our approximation facilitates the study of the global dynamics of the system and the exploration of effects of parameter variation on population dynamics. After verifying that the approximate model agrees qualitatively with both experiment and with the Keller-Segel model, we define a condition on model parameters that determines whether the bacterial populations will combine or turn around, then develop and numerically solve a boundary value problem to find the boundary between these two outcomes in various parameter spaces.

Our results leave us with predictions about the mechanisms by which the \emph{E. coli} populations manage to turn around and move away from each other and the nutrient gradient. Model (\ref{eq:gauss}) shows that the center of mass of a bacterial population is generically between the center of mass of its chemoattractant and that of the  {nutrient} early in the experiment. This relationship allows the bacteria to turn around if the outward attraction toward the chemoattractant becomes stronger than the inward attraction toward the nutrient.  Outward attraction can overcome inward attraction in a number of ways. For example, if the 
amount of nutrient between the bacterial populations is small, it is likely to yield a small attractive gradient and hence the bacteria will turn back toward the chemoattractant.  Our model predicts that if the medium in which the bacteria are suspended is too fluid,  then the two bacterial populations will likely combine, because the chemoattractant will spread across the spatial domain, removing the driver of the direction reversal. Variations in the total amount of available nutrient or fluidity of the medium can therefore lead to qualitative changes in the behavior of the bacteria.

The systematic predictions made by our approximate system agree qualitatively with specific simulations of the Keller-Segel model. Figs \ref{fig:PDECombine} and \ref{fig:PDETurn} provide particular examples in which decreasing the initial amount of nutrient can cause the bacterial populations to switch from a combination outcome to a turnaround outcome. Similarly, increasing the diffusivity of all three populations results in the bacterial populations combining (results not shown). This agreement suggests that our Gaussian approximation system offers reasonable predictions to be tested experimentally.

Our analysis of the two-population system assumes that both bacterial populations were of equal size. This assumption 
reduces the number of free parameters but might be unrealistic, as population size could vary between the two populations during an experiment. Simulations of system (\ref{eq:gauss2}) with unequal but similar population sizes agree qualitatively with those presented in this paper, and the various forms of dynamics we observed therefore do not result from a perfect symmetry in the populations.

Our approximated ODE system and the method used to derive it provide an efficient and tractable framework for analyzing the transient dynamics of complex systems. A similar analysis was conducted in \cite{Alt}, in which the authors used singular perturbation techniques to derive a Lotka-Volterra-like ODE competition model between invasive bacteria and host leukocytes from a Keller-Segel system adapted to model the inflammatory response due to bacterial infection. The resulting system allowed the authors to conduct an analysis of global behavior as a function of model parameters but removed all spatial aspects of the system. Our approximation preserves the spatial dimension by considering the temporal dynamics of the key quantities that characterize spatial features of our model populations.

There are several open directions related to this study. The first is to explore other, more quantitatively accurate approximations to the population distributions. While the bacteria maintains a Gaussian pulse-like distribution in the Keller-Segel model, the chemoattractant and nutrient populations do not necessarily do the same, especially as the two populations interact. One could impose a different assumption on the distribution of the chemical concentrations, the results of which could be important in understanding details of transient behaviors. 
Our study is primarily qualitative in flavor, and a more quantitatively accurate model could produce more precise experimental predictions. Our heuristic approximation could easily be applied as a method of moment closure for other spatiotemporal models whose nonlinearities make parameter exploration and transient analysis tedious or impossible.  {However, care must be taken when approximating a pulse by a Gaussian distribution. In \cite{Salman2} and \cite{Saragosti}, the bacterial pulses observed were asymmetrical, making a Gaussian a bad fit. In such cases, the spatial profiles can be approximated using parameterized non-Gaussian functions allowing one to derive ODE systems for profile parameters, but the added asymmetry would likely forfeit the analytical advantage of being able to compute the integrals in system (\ref{eq:gauss})}. Finally, it would be interesting to apply our Gaussian approximation method to a two-dimensional Keller-Segel model and explore transient dynamics, asymptotic states, and pattern formation.

\section*{Materials and Methods}
In this section we present our experimental methods, define our adaptation of a Keller-Segel model and derive a simplified model that captures important aspects of the collective bacterial motion in a framework that is more amenable to investigation.

\subsection*{Experimental methods}\label{sec:expMeth}

Wild type \emph{Escherichia coli} (\emph{E. coli}) RP437, expressing either yellow fluorescent protein (YFP) or red fluorescent protein (tdTomato) from a medium copy number plasmid (pZA) under the control of the constitutive  {$\lambda$-Pr promoter}, were grown in M9 minimal medium supplemented with 1g/l casamino acids, and 4g/l glucose (M9CG) at $30^\circ$C until early exponential growth phase (Optical Density at 600nm (OD$_{600\textrm{nm}})=0.1$) \cite{Lutz}. The cultures were then centrifuged for 5 minutes at 10,000 rpm, and resuspended in fresh M9CG medium at an OD$_{600\textrm{nm}}$=0.3. Each of the bacterial cultures was loaded onto one end of a set of $\sim$2cm long, thin channels (800 $\mu$m wide, 20-25 $\mu$m deep) fabricated in polydimethylsiloxane (PDMS) and adhered to a microscope glass slide (Fig \ref{fig:expSetUp}). The channels were pre-filled with fresh M9CG medium.  {The bacteria on both ends of the channel were allowed to migrate into the channel. The channel was then sealed on both ends with an epoxy glue to prevent any flow through the channel during observations. This method was successfully used in a previous study \cite{Salman2}, and the absence of a flow was confirmed by adding latex beads to the medium and observing their motion to verify that it is diffusive.} The sample was then mounted onto an inverted microscope (Zeiss Axiovert 40 CFL), and the bacteria were observed  {at room temperature ($\sim22^\circ$C)} in fluorescence mode using a 2.5x objective. Shortly after  {sealing the channel ends} ($\sim$10 - 20 minutes), a sharp accumulation peak appeared at each end of the channel, which then proceeded to advance as a pulse towards the center of the channel following a food gradient created by the bacterial food consumption at the densely populated ends (for more details about this phenomenon see for example \cite{Park,Salman2,Saragosti}).  {When the two pulses were ~4mm away from each other, their dynamics were recorded, each population} in its corresponding fluorescence colors, at a rate of 1 image/9 seconds using a charge-coupled device (CCD) camera (Progress MF, Jenoptik). The fluorescence profile reflecting the bacterial concentration along the channel was measured using ImageJ (NIH). For each of the examples presented in Fig \ref{fig:expResults}A-B, the fluorescence intensity is depicted in units of the maximal measured fluorescence at the peak of the concentration and the background was subtracted for better comparison.  {Note that due to technical limitations of imaging that prevent us from acquiring larger images, Fig \ref{fig:expResults} A1 and B1 display only a $\sim3.5$mm long section of the channel where the two populations meet. Therefore, the integral of the fluorescence intensity profile over the whole range appears to be changing, but this is not due to bacterial reproduction. Under the experimental conditions used here (culture medium being M9CG and incubation temperature being $\sim 22^\circ$C) the reproduction rate of the bacteria is about once every 2 hours.}

\subsection*{Keller-Segel model  {framework}}\label{sec:KS}

We denote by $b(t,x)$ the bacterial concentration at time $t$ and spatial coordinate $0\leq x\leq L$. The cell density moves both by linear diffusion and by chemotaxis up a chemical gradient. A full derivation of the differential equations modeling such dynamics can be found in, for example, \cite{Hillen,Horstmann}. Here we consider the effects of two chemical densities: a chemoattractant (glycine; {see Supporting Information}) produced by the bacteria, $a(t,x)$, and an externally added nutrient, $\phi(t,x)$. We assume that the chemoattractant is produced by the bacteria at constant rate $r$ and naturally degrades at rate $\delta$ and that the bacteria consume the nutrient at constant rate $\kappa$.  {The replication time of the bacteria under the experimental conditions was measured to be about 2 hours. On the other hand, all of the experiments lasted about an hour, with an observation and measurement window of $\sim 20$ minutes. We therefore neglect the replication of the bacteria in our model.} Under these assumptions, the model we study is system (\ref{eq:KS}) (previously stated under Results):

\begin{equation}\tag{1}
\begin{aligned}
\frac{\partial b}{\partial t}&=D_b\frac{\partial^2 b}{\partial x^2}-\chi_a\frac{\partial}{\partial x}\left[b\frac{\partial a}{\partial x} \right]-\chi_\phi\frac{\partial}{\partial x}\left[b\frac{\partial\phi}{\partial x} \right]\\
\frac{\partial a}{\partial t}&=D_a\frac{\partial^2 a}{\partial x^2}+rb-\delta a\\
\frac{\partial \phi}{\partial t} &=D_\phi\frac{\partial^2 \phi}{\partial x^2}-\kappa b\phi
\end{aligned}
\end{equation}
with boundary conditions 
\begin{equation}\label{eq:BCs}
\left.\frac{\partial b}{\partial x}\right|_{x=0,L}=\left.\frac{\partial a}{\partial x}\right|_{x=0,L}=\left.\frac{\partial \phi}{\partial x}\right|_{x=0,L}=0.
\end{equation}

We impose minimal biological assumptions on our model: we ignore any effects of cell physiology on chemical sensing, such as signal-dependent sensitivity, and any cell kinetics. 
We note that the receptor-binding adaptation to the Keller-Segel model  (model (M2a) in \cite{Hillen}) produces similar qualitative results as we present in this paper, though we do not present those results here.

For the purpose of differentiating between two populations of bacteria in numerical simulations, we include in our model two identical populations of bacteria, $b$ and $\beta$, that each produce the same chemoattractant with spatiotemporal profiles, $a_1$ and $a_2$, respectively. These variables are differentiated by their initial distributions but their behaviors and influences on each other are otherwise identical.
 {With both populations included, the Keller-Segel model becomes system (\ref{eq:KS2})}
with boundary conditions 
\begin{equation}\label{eq:BCs2}
\left.\frac{\partial b_{1,2}}{\partial x}\right|_{x=0,L}=\left.\frac{\partial a_{1,2}}{\partial x}\right|_{x=0,L}=\left.\frac{\partial \phi}{\partial x}\right|_{x=0,L}=0.
\end{equation}
Importantly, the dynamics of the two-population model is identical to that of the single-population model because the system is linear in $b$ and $a$. That is, if $\tilde b=b+\beta$ and  {$\tilde a=a+\alpha$}, then the system of differential equations governing the dynamics of $\tilde b$,  {$\tilde a$}, and $\phi$ is exactly system (\ref{eq:KS}).

The initial profile of the externally added nutrient is given by the reflected sigmoid 

\begin{equation}\label{eq:initFood}
\phi(0,x)=\left\{\begin{array}{cc}
\phi_0/(1+\exp(-100x+10)), & 0\leq x\leq L/2,\\
\phi_0/(1+\exp(100x-90)), &    {L/2}< x\leq L
\end{array}\right.,
\end{equation}
where $\phi_0$ is a parameter.

We nondimensionalize model (\ref{eq:KS2}) as follows:
$$b=N\tilde b;\ \beta=N\tilde \beta;\ a_1=K\tilde a_1;\ a_1=K\tilde a_1;\ \phi=M\tilde \phi;\ x=\tilde x/L,$$
where $L$ is the domain length, and $N,$ $K,$ and $M$ are large numbers 
representing the maximum size of the bacterial, chemoattractant, and nutrient populations, respectively. Here we assume that  {$L=20$} cm, $K=M\sim 10^{14}$ and $N\sim10^8$. After nondimensionalization, the parameter values we use are those given in Table \ref{tab:params}. The natural dimensions (before nondimensionalization) are included. After nondimensionalization, all parameters have units $s^{-1}$, the spatial domain is the unit interval $[0,1]$. The boundary conditions (\ref{eq:BCs}) for the one bacterial population system (\ref{eq:KS}) therefore become
\begin{equation}\label{eq:BCsND}
\left.\frac{\partial b}{\partial x}\right|_{x=0,1}=\left.\frac{\partial a}{\partial x}\right|_{x=0,1}=\left.\frac{\partial \phi}{\partial x}\right|_{x=0,1}=0
\end{equation}
and the boundary conditions (\ref{eq:BCs2}) for the two bacterial population system (\ref{eq:KS2}) become
\begin{equation}\label{eq:BCs2ND}
\left.\frac{\partial b_{1,2}}{\partial x}\right|_{x=0,1}=\left.\frac{\partial a_{1,2}}{\partial x}\right|_{x=0,1}=\left.\frac{\partial \phi}{\partial x}\right|_{x=0,1}=0.
\end{equation}
For simplicity, we immediately replace the nondimensionalized symbols $\tilde b$, $\tilde \beta$, $\tilde a_1$, $\tilde a_2$, $\tilde \phi$, and $\tilde x$ with $b$, $\beta$, $a_1$, $a_2$ $\phi$, and $x$, respectively, in the nondimensionalized system.

Each of the model parameters except $\kappa$ are based on those found in \cite{Salman2} and \cite{Saragosti}. The model found in \cite{Salman2} assumes that nutrient consumption is proportional to the bacterial density only; that is $-\kappa b$. Here, nutrient consumption is given by a mass action term proportional to the product of the bacterial density and nutrient, $-\kappa b\phi$. The consumption rate used here is therefore much smaller than that found in the literature.

 \begin{center}
\begin{table}[H]
\begin{tabular}{ p{.25in}  p{2.5in}  p{.75in}  p{1.3in} }
\hline
\multicolumn{2}{c}{Parameter} & Value & Natural dimensions\\
\hline\hline \noalign{\smallskip}
$D_b$ & Diffusivity of bacteria & 0.001& cm$^2 $$\cdot$s$^{-1}$\\
$D_a$ & Diffusivity of attractant & 0.03 & cm$^2\cdot$s$^{-1}$\\
$D_\phi$ & Diffusivity of nutrient & 0.03 & cm$^2\cdot$s$^{-1}$\\
$\chi_a$ & Chemotactic sensitivity to attractant & 0.025 & cm$^3\cdot$s$^{-1}$\\
$\chi_\phi$ & Chemotactic sensitivity to nutrient & 0.015 & cm$^3\cdot$s$^{-1}$\\ 
$r$ & Production rate of attractant by bacteria & 0.05 & bacterium$^{-1}\cdot$ s$^{-1}$\\
$\delta$ & Natural decay rate of attractant & 0.005 & s$^{-1}$\\
$\kappa$ & Consumption rate of nutrient by bacteria & 0.001 & (bacterium/cm)$^{-1}\cdot$s$^{-1}$ \\ 
\end{tabular}
\caption{
Parameters used in model (\ref{eq:KS}). Before nondimensionalization, all parameters other than $\kappa$ are based on those found in \cite{Salman2} and \cite{Saragosti}.}
\label{tab:params}
\end{table}
\end{center}


\subsection*{Gaussian Moment Closure}\label{sec:gauss}

Here we approximate the temporal dynamics of the spatial moments of the bacterial, chemoattractant, and nutrient distributions. Because the pulse-pulse interaction occurs in the interior of the spatial domain, we are concerned with transient behavior of the bacteria away from the spatial boundaries. We therefore remove boundary effects by considering the spatial domain $(-\infty,\infty)$.

The $i^{th}$ moment of population $s$, $s\in\{b,a,\phi\}$, is defined as
\begin{equation*}
s_i(t)=\displaystyle \int_{-\infty}^{\infty} x^is(t,x)dx,
\end{equation*}
for $i=0,1,\dots$
For each of the three populations $s$, the quantity $s_0$ is interpreted as the total amount of $s$ in the system, $\mu_s:=s_1/s_0$ is the location of the center of mass of $s$, and $\sigma^2_s:=s_2/s_0-\left(s_1/s_0\right)^2$ is the variance of the profile of $s$.
Differentiating $s_0$, $\mu_s$, and $\sigma_s^2$ for $s\in\{b,a,\phi\}$, with respect to time provides a system of differential equations describing the temporal dynamics of these variables. For example, the differential equation governing the dynamics of $b_0$ is

\begin{equation}\label{eq:b0}
\begin{aligned}
 \dot b_0 &= \displaystyle \int_{-\infty}^\infty \frac{\partial }{\partial t}\left[b(t,x)\right] dx\\
 &= \displaystyle \int_{-\infty}^\infty \left(D_b\frac{\partial^2 b}{\partial x^2}-\chi_a\frac{\partial}{\partial x}\left[b\frac{\partial a}{\partial x} \right]-\chi_\phi\frac{\partial}{\partial x}\left[b\frac{\partial\phi}{\partial x} \right]\right)dx\\
 &=\left.\left(D_b\frac{\partial b}{\partial x}-\chi_ab\frac{\partial a}{\partial x} -\chi_\phi b\frac{\partial\phi}{\partial x}\right)\right|_{-\infty}^{\infty}\\
 &=0
 \end{aligned}
\end{equation}
and that of $\mu_b$ is
\begin{equation*}
\begin{aligned}
\dot \mu_b&=\frac{d}{dt}\left[b_1/b_0\right]\\
&=\frac{1}{b_0}\displaystyle \int_{-\infty}^\infty \frac{\partial }{\partial t}\left[xb(t,x)\right] dx\\
&= \frac{1}{b_0}\displaystyle \int_{-\infty}^\infty x\left(D_b\frac{\partial^2 b}{\partial x^2}-\chi_a\frac{\partial}{\partial x}\left[b\frac{\partial a}{\partial x} \right]-\chi_\phi\frac{\partial}{\partial x}\left[b\frac{\partial\phi}{\partial x} \right]\right)dx\\
&= \frac{1}{b_0}\left[x\left(D_b\frac{\partial b}{\partial x}-\chi_ab\frac{\partial a}{\partial x} -\chi_\phi b\frac{\partial\phi}{\partial x}\right)\Bigr|_{-\infty}^{\infty}-\int_{-\infty}^\infty (D_b\frac{\partial b}{\partial x}-\chi_ab\frac{\partial a}{\partial x} -\chi_\phi b\frac{\partial\phi}{\partial x}) dx\right]\\
&=\frac{1}{b_0}\left[0-D_bb\bigr|_{-\infty}^\infty+\chi_a\langle ba_x\rangle+\chi_\phi\langle b\phi_x\rangle\right]\\
&=\frac{\chi_a\langle ba_x\rangle+\chi_\phi\langle b\phi_x\rangle}{b_0},
\end{aligned}
\end{equation*}
where $1/b_0$ factors out because $b_0$ is a constant by (\ref{eq:b0}), the boundary terms are zero by assumption, and $$\langle fg\rangle=\displaystyle \int_{-\infty}^\infty f(x)g(x)dx.$$
Similarly, differentiating each of the seven remaining variables produces the following system of nine ordinary differential equations:

\begin{equation}\label{eq:gauss}
\begin{aligned}
\dot{b}_0&=0\\
\dot{a}_0&=rb_0-\delta a_0\\
\dot{\phi}_0&=-\kappa\langle b\phi\rangle\\
\dot{\mu}_b&=\frac{\chi_a\langle ba_x\rangle+\chi_\phi\langle b\phi_x\rangle}{b_0}\\
\dot{\mu}_a&=\frac{rb_0}{a_0}(\mu_b-\mu_a)\\
\dot{\mu}_\phi&=-\frac{\kappa}{\phi_0}(\langle xb\phi\rangle-\mu_\phi\langle b\phi\rangle)\\
\dot{\sigma}^2_b&=2D_b+2\frac{\chi_a\langle(x-\mu_b)ba_x\rangle+\chi_\phi\langle(x-\mu_b)b\phi_x\rangle}{b_0}\\
\dot{\sigma}^2_a&=2D_a+\frac{rb_0}{a_0}\left(\sigma^2_b-\sigma^2_a+(\mu_b-\mu_a)^2\right)\\
\dot{\sigma}^2_\phi&=2D_\phi+\frac{\kappa}{\phi_0}\left(\sigma^2_\phi\langle b\phi\rangle-\langle(x-\mu_\phi)^2b\phi \rangle\right).
\end{aligned}
\end{equation}
Unless otherwise specified, the parameter values used are those found in Table \ref{tab:GaussParams}. These values are based on those used in the Keller-Segel model (\ref{eq:KS}), though the chemotactic sensitivity and diffusivity parameters are affected by the increased spatial scale.

 \begin{center}
\begin{table}[H]
\begin{tabular}{l  p{2.6in} l p{1.25in}  p{1.3in} }
\hline
\multicolumn{2}{c}{Variable} & Initial condition\\
\hline\hline \noalign{\smallskip}
\smallskip
$b_0$ & Total bacteria & 3\\ \smallskip
$a_0$ & Total chemoattractant & $rb_0/\delta$\\ \smallskip
$\phi_0$ & Total nutrient & 35\\ \smallskip
$\mu_b$ & Center of mass of bacteria & 0\\ \smallskip
$\mu_a$ & Center of mass of chemoattractant & 0\\ \smallskip
$\mu_\phi$ & Center of mass of nutrient & 0.5\\ \smallskip
$\sigma_b^2$ & Variance of the bacteria profile & 0.005\\ \smallskip
$\sigma_a^2$ & Variance of the chemoattractant profile & 0.2\\ 
$\sigma_\phi^2$ & Variance of the nutrient profile & 0.1\\
\hline
\multicolumn{2}{c}{Parameter} & Value & Natural dimension\\
\hline\hline \noalign{\smallskip}
\smallskip
$D_b$ & Diffusivity of bacteria & $10^{-5}$& cm$^2 $$\cdot$s$^{-1}$\\ \smallskip
$D_a$ & Diffusivity of attractant & 0.0002& cm$^2 $$\cdot$s$^{-1}$\\ \smallskip
$D_\phi$ & Diffusivity of nutrient & 0.0002& cm$^2 $$\cdot$s$^{-1}$\\ \smallskip
$\chi_a$ & Chemotactic sensitivity to attractant & 0.00025& cm$^3\cdot$s$^{-1}$\\ \smallskip
$\chi_\phi$ & Chemotactic sensitivity to nutrient & 0.0002& cm$^3\cdot$s$^{-1}$\\ \smallskip
$r$ & Production rate of attractant by bacteria & 0.05& bacterium$^{-1}\cdot$s$^{-1}$\\ \smallskip
$\delta$ & Natural decay rate of attractant & 0.005& s$^{-1}$\\ \smallskip
$\kappa$ & Consumption rate of nutrient by bacteria & 0.001& (bacterium/cm)$^{-1}\cdot$s$^{-1}$\\ 
\end{tabular}
\caption{System variables and parameters  (dimensionless) used in model (\ref{eq:gauss}).}
\label{tab:GaussParams}
\end{table}
\end{center}

We require a method of moment closure to evaluate the mixed moments that appear in system (\ref{eq:gauss}). Moreover, as we are primarily concerned with the interaction of bacterial pulses, we choose an approximation to the spatial distribution of each population that preserves a pulse-like shape. To this end, we assume that each population behaves like a Gaussian distribution; that is,
\begin{equation}\label{eq:gaussian}
s(t,x)=\frac{s_0}{\sigma_s\sqrt{\pi}}\exp\left(\frac{-(x-\mu_s)^2}{\sigma_s^2}\right).
\end{equation}
We note that there is nothing particularly special about the Gaussian distribution, other than the empirical observation that the spatial profile of each population looks approximately Gaussian. We could similarly have chosen any function that approximates a pulse, such as $\sech(x)$, but the Gaussian function allows us to explicitly evaluate each integral that appears in system (\ref{eq:gauss}).


\subsubsection*{Stability of uniform state}\label{sec:gaussStab}
While we cannot expect perfect quantitative agreement between models (\ref{eq:KS}) and (\ref{eq:gauss}), we can confirm that (\ref{eq:gauss}) reproduces key qualitative behaviors of (\ref{eq:KS}). We first consider the linear stability of all equilibrium points of system (\ref{eq:gauss}). After substituting (\ref{eq:gaussian}) into system (\ref{eq:gauss}), the $\phi$-subsystem becomes
\begin{equation*}
\begin{aligned}
\dot{\phi}_0&=\frac{-\kappa b_0\phi_0}{\sqrt{\pi}(\sigma_b^2+\sigma_\phi^2)^{1/2}}\exp\left(\frac{-(\mu_b-\mu_\phi)^2}{\sigma_b^2+\sigma_\phi^2}\right)\\
\dot{\mu}_\phi&=\frac{-\kappa b_0(\mu_b-\mu_\phi)\sigma_\phi^2}{\sqrt{\pi}(\sigma_b^2+\sigma_\phi^2)^{3/2}}\exp\left(\frac{-(\mu_b-\mu_\phi)^2}{\sigma_b^2+\sigma_\phi^2}\right)\\
\dot{\sigma}^2_\phi&=2D_\phi+\frac{\kappa b_0\sigma_\phi^2\left[(\sigma_b^2+\sigma_\phi^2)(\sigma_b^2+2\sigma_\phi^2)-2(\mu_b-\mu_\phi)^2\sigma_\phi^2\right]}{2\sqrt{\pi}(\sigma_b^2+\sigma_\phi^2)^{5/2}}\exp\left(\frac{-(\mu_b-\mu_\phi)^2}{\sigma_b^2+\sigma_\phi^2}\right).
\end{aligned}
\end{equation*}
The differential equation for $\phi_0$ readily implies that $\phi_0\to0$ as $t\to\infty$, and the differential equation for $\sigma^2_\phi$ is initially positive and increasing in $\sigma^2_\phi$, implying that $\sigma_\phi^2\to\infty$ as $t\to\infty$. For large time, the nutrient population therefore necessarily becomes consumed entirely and diffuses out to the uniform state besides, and will thereby become uncoupled from the $b$-$a$-subsystem. Consequently, the linear stability of any equilibrium point in the $b$-$a$-subsystem is unaffected by the $\phi$-subsystem with respect to perturbations that do not involve the nutrient $\phi$. Any perturbation involving $\phi$ will only affect the location of the center of mass of $b$ and $a$, but not the width of their distributions.

The differential equation for $\mu_a$ indicates that we must have $\mu_b=\mu_a$ at any equilibrium point, but the specific value of these two variables is arbitrary (in other words, the bacterial and chemical pulse must accumulate around the same spatial coordinate, but that coordinate can be anywhere). We therefore introduce the relative coordinate $\mu=\mu_b-\mu_a$.  Under this transformation, imposing assumption (\ref{eq:gaussian}) and evaluating the integrals remaining in the $b$-$a$-subsystem in (\ref{eq:gauss}) produces

\begin{equation}\label{eq:gaussAB2}
\begin{aligned}
\dot{b}_0&=0\\
\dot{a}_0&=rb_0-\delta a_0\\
\dot{\mu}&=\left[\frac{-2\chi_aa_0}{\sqrt{\pi}(\sigma_b^2+\sigma_a^2)^{3/2}}\exp\left(\frac{-\mu^2}{\sigma_b^2+\sigma_a^2}\right)-\frac{rb_0}{a_0}\right]\mu\\
\dot{\sigma}^2_b&=2D_b-2\frac{\chi_aa_0\sigma_b^2}{\sqrt{\pi}(\sigma_b^2+\sigma_a^2)^{5/2}}\left(\sigma_b^2+\sigma_a^2-2\mu^2\right)\exp\left(\frac{-\mu^2}{\sigma_b^2+\sigma_a^2}\right)\\
\dot{\sigma}^2_a&=2D_a+\frac{rb_0}{a_0}(\sigma^2_b-\sigma^2_a+\mu^2).
\end{aligned}
\end{equation}

From the first two equations, any fixed point of this system must satisfy $b_0^*=constant$ and $a_0^*=rb_0^*/\delta$. Since the term inside the brackets in the $\mu$ equation is strictly negative, any fixed point must also satisfy $\mu=0$. The remaining two-variable system is

\begin{equation}\label{eq:ABvar}
\begin{aligned}
\dot{\sigma}^2_b&=2\left(D_b-\frac{\chi_arb_0^*\sigma_b^2}{\sqrt{\pi}\delta(\sigma_b^2+\sigma_a^2)^{3/2}}\right)\\
\dot{\sigma}^2_a&=2D_a+\delta(\sigma^2_b-\sigma^2_a).
\end{aligned}
\end{equation} 

The generic cases of the nullclines for system (\ref{eq:ABvar}) are plotted in Fig \ref{fig:nullclines1}. In Fig \ref{fig:nullclines1}A, the total amount of bacteria is $b_0=0.01$ and the system contains no fixed points. The variance of both populations blows up to infinity as time gets large for any initial condition; that is, the bacterial population will always diffuse out into a uniform state if the population size is too low. In Fig \ref{fig:nullclines1}B, $b_0$ is increased to $0.012$, and two fixed points exist: a stable node and a saddle point. The stable equilibrium point is analogous to the pulse solution of system (\ref{eq:KS}): the bacterial population and its chemoattractant  accumulate around the same center of mass ($\mu=\mu_b-\mu_a=0$) with a small variance around this point. Starting with a variance in the bacterial population that is too large (that is, to the right of the separatrix of the saddle point), however, results in the variance of both populations increasing without bound. This case is analogous to the system converging to the uniform solution, and so model (\ref{eq:gauss}) is generically bistable when $b_0$ is above a critical threshold.

Fig \ref{fig:varBif1} shows the saddle-node bifurcation as a function of $b_0$, the total amount of bacteria. We can explicitly calculate the critical value of $b_0$ at which the bifurcation occurs as a function of model parameters. The nullclines of system (\ref{eq:ABvar}) intersect when

%
%
%

\begin{equation*}
D_b-\frac{\chi_arb_0\sigma_b^2}{\sqrt{\pi}\delta(2\sigma_b^2+2D_a/\delta)^{3/2}}=0,
\end{equation*}
or equivalently,
\begin{equation}\label{eq:intersection}
(\sigma_b^2)^3+\left(3\frac{D_a}{\delta}-\frac{\chi_a^2r^2b_0^2}{8\pi\delta^2D_b^2}\right)(\sigma_b^2)^2+\frac{3D_a^2}{\delta^2}\sigma_b^2+\frac{D_a^3}{\delta^3}=0.
\end{equation}
When 
\begin{equation}\label{eq:critBif}
3\frac{D_a}{\delta}-\frac{\chi_a^2r^2b_0^2}{8\pi\delta^2D_b^2}=-\frac{15}{4}\frac{D_a}{\delta},
\end{equation}
equation (\ref{eq:intersection}) can be written $$\left(\sigma_b^2-2\frac{D_a}{\delta}\right)^2\left(\sigma_b^2+\frac{D_a}{4\delta}\right)=0,$$ and therefore equation (\ref{eq:critBif}) is the condition for when the two positive roots of (\ref{eq:intersection}) coalesce. This condition gives us the critical bifurcation value for $b_0$, 
\begin{equation}
b_0^*=\frac{D_b\sqrt{54\pi D_a\delta}}{r\chi_a }.
\end{equation}
If $b_0>b_0^*$, then equation (\ref{eq:intersection}) has two roots and a stable pulse solution of system (\ref{eq:gaussAB2}) exists, and if $b_0<b_0^*$, then the equation has no roots and the uniform state is the only asymptotic solution of the system.
Comparison to the critical value of $b^{tot}$ in the Keller-Segel system (\ref{eq:KS}), $$b^{tot}=\frac{D_b(\pi^2D_a+\delta)}{r\chi_a},$$ shows that a change in any of the model parameters  for (\ref{eq:gauss}) produces the same qualitative effect on the threshold as in the PDE system (\ref{eq:KS}).

\subsubsection*{Capturing turnaround}\label{sec:gaussTurn}

For our numerical simulations of system (\ref{eq:gauss}), we introduce a second bacterial population $\beta$ and corresponding chemoattractant concentration $\alpha$, both of which we again assume maintain a Gaussian profile. We assume the chemoattractant produced by this new population is the same chemical produced by the original bacterial population, and with the addition of these variables, system (\ref{eq:gauss}) becomes

{
\begin{equation}\label{eq:gauss2}
\begin{aligned}
\dot{b}_0&=0\\
\dot{a}_0&=rb_0-\delta a_0\\
\dot{\beta}_0&=0\\
\dot{\alpha}_0&=r\beta_0-\delta \alpha_0\\
\dot{\phi}_0&=-\kappa\langle b\phi\rangle-\kappa\langle \beta\phi\rangle\\
\dot{\mu}_b&=\frac{\chi_a\langle ba_x\rangle+\chi_a\langle b\alpha_x\rangle+\chi_\phi\langle b\phi_x\rangle}{b_0}\\
\dot{\mu}_a&=\frac{rb_0}{a_0}(\mu_b-\mu_a)\\
\dot{\mu}_\beta&=\frac{\chi_a\langle \beta a_x\rangle+\chi_a\langle \beta \alpha_x\rangle+\chi_\phi\langle \beta \phi_x\rangle}{\beta_0}\\
\dot{\mu}_\alpha&=\frac{r\beta_0}{\alpha_0}(\mu_\beta-\mu_\alpha)\\
\dot{\mu}_\phi&=-\frac{\kappa}{\phi_0}(\langle xb\phi\rangle-\mu_\phi\langle b\phi\rangle)-\frac{\kappa}{\phi_0}(\langle x\beta \phi\rangle-\mu_\phi\langle \beta \phi\rangle)\\
\dot{\sigma}^2_b&=2D_b+2\frac{\chi_a\langle(x-\mu_b)ba_x\rangle+\chi_a\langle(x-\mu_b)b\alpha_x\rangle+\chi_\phi\langle(x-\mu_b)b\phi_x\rangle}{b_0}\\
\dot{\sigma}^2_a&=2D_a+\frac{rb_0}{a_0}\left(\sigma^2_b-\sigma^2_a+(\mu_b-\mu_a)^2\right)\\
\dot{\sigma}^2_\beta&=2D_b+2\frac{\chi_a\langle(x-\mu_\beta)\beta a_x\rangle+\chi_a\langle(x-\mu_\beta)\beta \alpha_x\rangle+\chi_\phi\langle(x-\mu_\beta)\beta \phi_x\rangle}{\beta_0}\\
\dot{\sigma}^2_\alpha&=2D_a+\frac{r\beta_0}{\alpha_0}\left(\sigma^2_\beta-\sigma^2_\alpha+(\mu_\beta-\mu_\alpha)^2\right)\\
\dot{\sigma}^2_\phi&=2D_\phi+\frac{\kappa}{\phi_0}\left(\sigma^2_\phi\langle b\phi\rangle+\sigma^2_\phi\langle \beta\phi\rangle-\langle(x-\mu_\phi)^2b\phi \rangle-\langle(x-\mu_\phi)^2\beta\phi \rangle\right).
\end{aligned}
\end{equation}
}

We will first show that any equilibrium point of system (\ref{eq:gauss2}) must satisfy $\mu_b=\mu_\beta=\mu_a=\mu_\alpha$. From the first four equations of the system, we have that $a_0^*=rb_0^*/\delta$ and $\alpha_0^*=r\beta_0^*/\delta$ at any equilibrium point, where $b_0^*$ and $\beta_0^*$ are constants. The $\mu_a$ and $\mu_\alpha$ equations require that $\mu_b=\mu_a$, and $\mu_\beta=\mu_\alpha$, respectively. Moreover, as in system (\ref{eq:gauss}), the  {nutrient} population is entirely transient, $\phi_0\to0$ as $t\to\infty$, so the nutrient will not affect asymptotic stability. Next, we introduce the relative center of mass coordinate $\mu_{b\beta}=\mu_b-\mu_\beta$. Imposing $\mu_b=\mu_a$ and $\mu_\beta=\mu_\alpha$, the differential equation governing $\mu_{b\beta}$ is

\begin{equation}\label{eq:relPos}
{
\dot{\mu}_{b\beta}=
\left(-\frac{2\chi_ar\beta_0}{\delta\sqrt{\pi}(\sigma_b^2+\sigma_\alpha^2)^{3/2}}\exp\left(\frac{-\mu_{b\beta}^2}{\sigma_b^2+\sigma_\alpha^2}\right)-\frac{2\chi_arb_0}{\delta\sqrt{\pi}(\sigma_\beta^2+\sigma_a^2)^{3/2}}\exp\left(\frac{-\mu_{b\beta}^2}{\sigma_\beta^2+\sigma_a^2}\right)\right)\mu_{b\beta}.}
\end{equation}The expression multiplying $\mu_{b\beta}$ is strictly negative, and so $\dot{\mu}_{b\beta}=0$ if and only if $\mu_{b\beta}=0$. We have therefore shown that, at steady state, $\mu_a=\mu_b=\mu_\beta=\mu_\alpha$.

The remaining dynamical variables are governed by the system

\begin{equation}\label{eq:gaussTwoPopAB}
{
\begin{aligned}
\dot{\sigma}^2_b&=2D_b-2\frac{\chi_arb_0\sigma_b^2}{\delta\sqrt{\pi}(\sigma_b^2+\sigma_a^2)^{5/2}}\left(\sigma_b^2+\sigma_a^2\right)-2\frac{\chi_ar\beta_0\sigma_b^2}{\delta\sqrt{\pi}(\sigma_b^2+\sigma_\alpha^2)^{5/2}}\left(\sigma_b^2+\sigma_\alpha^2\right)\\
\dot{\sigma}^2_a&=2D_a+\delta(\sigma^2_b-\sigma^2_a)\\
\dot{\sigma}^2_\beta&=2D_b-2\frac{\chi_arb_0\sigma_\beta^2}{\delta\sqrt{\pi}(\sigma_\beta^2+\sigma_a^2)^{5/2}}\left(\sigma_\beta^2+\sigma_a^2\right)-2\frac{\chi_ar\beta_0\sigma_\beta^2}{\delta\sqrt{\pi}(\sigma_\beta^2+\sigma_\alpha^2)^{5/2}}\left(\sigma_\beta^2+\sigma_\alpha^2\right)\\
\dot{\sigma}^2_\alpha&=2D_a+\delta(\sigma^2_\beta-\sigma^2_\alpha).
\end{aligned}}
\end{equation}


If we make the additional simplifying assumptions that the two bacterial populations are of the same size (that is, $b_0=\beta_0$) and have the same initial variance, and the two chemoattractant populations have the same initial variance, then the first two equations are identical to the second two equations in system (\ref{eq:gaussTwoPopAB}), and consequently $\sigma_b^2=\sigma_\beta^2$ for all time (these assumptions ease analysis but are not necessary to achieve the results presented below; see Discussion). Imposing these conditions, system (\ref{eq:gaussTwoPopAB}) reduces to the two-dimensional system

\begin{equation}\label{eq:gaussAB2}
\begin{aligned}
\dot{\sigma}^2_b&=2\left(D_b-2\frac{\chi_arb_0\sigma_b^2}{\delta\sqrt{\pi}(\sigma_b^2+\sigma_a^2)^{3/2}}\right)\\
\dot{\sigma}^2_a&=2D_a+\delta(\sigma^2_b-\sigma^2_a).
\end{aligned}
\end{equation}
This system is nearly identical to system (\ref{eq:ABvar}), the only difference being that the $\chi_arb_0\sigma_b^2/(\delta\sqrt{\pi}(\sigma_b^2+\sigma_a^2)^{3/2})$ term is doubled in (\ref{eq:gaussAB2}) because there are now two bacterial populations producing chemoattractant. System (\ref{eq:ABvar}) therefore produces the same saddle-node bifurcation structure as the one-population system (\ref{eq:gauss}). We note that the same bifurcation will occur if the two populations are not of equal size, though the mathematical details become tedious and no more informative than in this simplified case.

\begin{figure}[H]
{\centering
\includegraphics[width=3in]{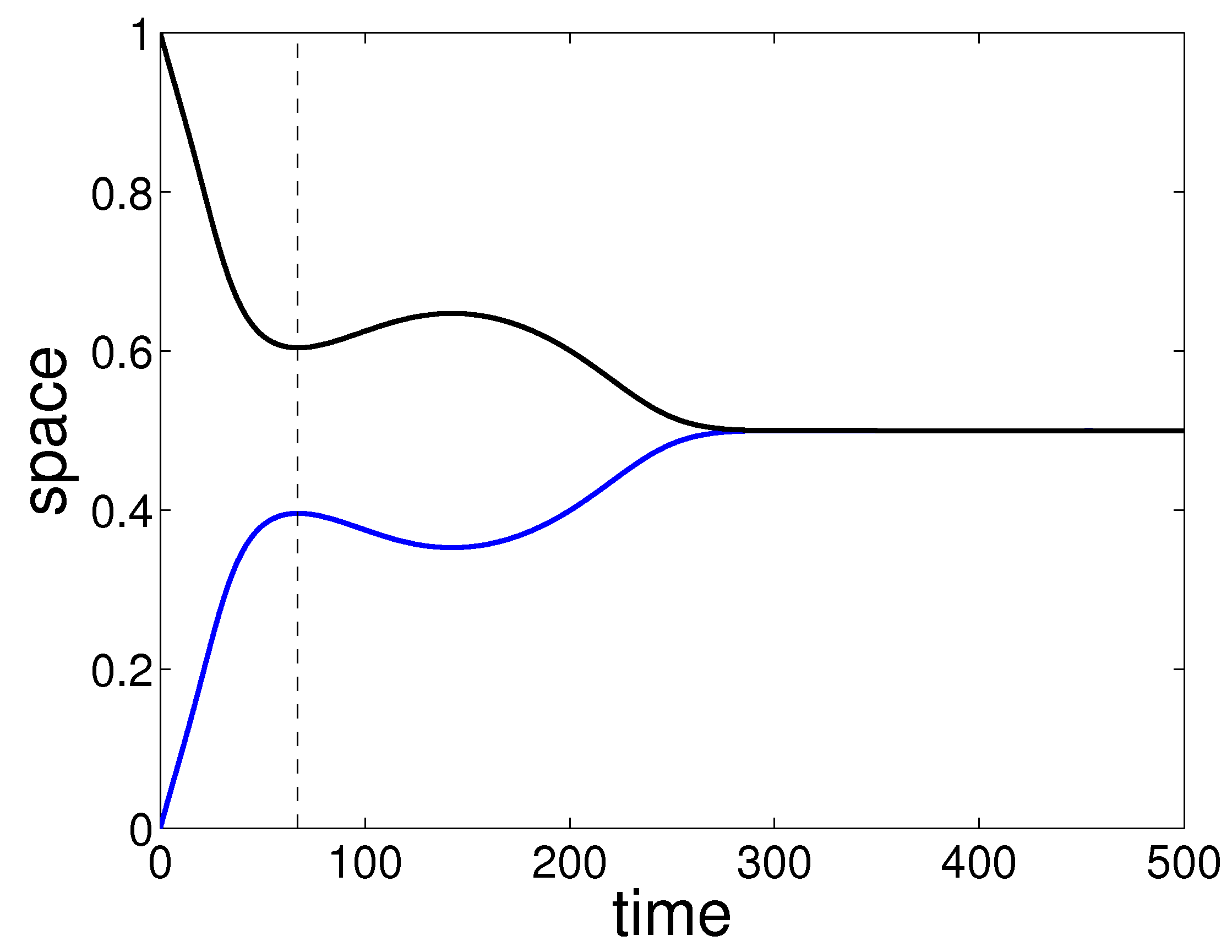}

}
\caption[False turnaround.]{False turnaround. The vertical dashed line marks a turnaround in the center of mass of both populations, but the populations combine together a short time later.}
\label{fig:combineBAD}
\end{figure}

Our goal in studying system (\ref{eq:gauss2}) is to determine parameter conditions under which the two populations combine and those under which they turn around. Because equilibria require that $\mu_b=\mu_\beta$, the two populations will necessarily combine in asymptotic time, in contrast to the Keller-Segel model (\ref{eq:KS2}). We therefore must take care in deciding what qualifies as a turnaround in model (\ref{eq:gauss2}). One possible condition is that $\dot{\mu}_b(t_1)=0$ and $\dot{\mu}_\beta(t_2)=0$ for some times $t_1$ and $t_2$ (indicating that the centers of mass of both populations have changed direction). However, this condition is not sufficient to determine when the populations turn around and move away from one another. Fig \ref{fig:combineBAD} shows an example where both populations quickly turn around but shortly thereafter turn back around and combine. Though the center of mass of each population does change direction in this example, the overall outcome is not compatible with experimentally observed turnaround, in which the two populations accumulate along opposite ends of the domain. We therefore adopt the following more robust definition of turnaround.

The chemotactic attraction of a bacterial population decays exponentially with the distance between the center of mass of the bacteria and that of the chemoattractant (see equation (\ref{eq:relPos})). Thus, if the two bacterial populations are sufficiently far apart, then the chemotactic pull from each pulse of chemoattractant to the more distant bacterial population is negligible, and the populations can separately approach a meta-stable state: each population develops into its own pulse-like structure, subject to only an exponentially small effect from the other population's chemoattractant. This state is intuitively consistent with the experimental state in which two bacterial populations accumulate along the boundaries of the domain. In asymptotic time, the two populations will always combine, but the farther apart the two populations are, the longer it will take for the combination to occur. Once the populations do become sufficiently close, however, the relative effect that each population experiences from the other population's chemoattractant becomes nontrivial, and they combine together relatively quickly. We hence reason that if the two populations have not combined after a large but finite amount of time, they must be in a meta-stable non-combined state. We therefore take as our condition for turnaround that the centers of mass of the two populations are distinct after a large amount of time; that is, that $|\mu_b(t_c)-\mu_\beta(t_c)|>\epsilon$ for some small, fixed distance $\epsilon$ at some large time $t=t_{c}$. 

\begin{figure}[H]
{\centering
\includegraphics{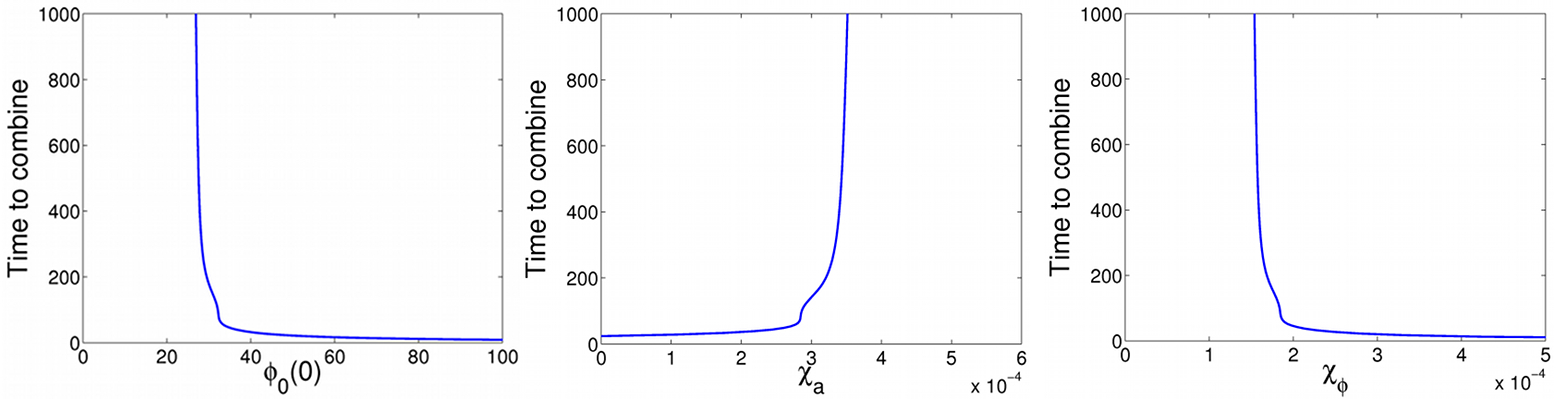}

}
\caption[Dependence of time to combine on model parameters.]{Dependence of time to combine on model parameters. In order to combine at time t, the parameter on the horizontal axis must be the value specified by the curve. Sensitivity of the time to combine, $t_{c}$, on each parameter considered decreases once $t_c$ exceeds some quantitative threshold. Similar figures for parameters $N_0$ and $D$ not shown.}
\label{fig:combineTime}
\end{figure}

We must take care in choosing values for $\epsilon$ and $t_c$. For instance, in order to establish a boundary between the turnaround outcome and the combination outcome, we must choose $\epsilon$ small enough so that the two populations will quickly combine if their center of masses are $\epsilon$ apart. We determine through numerical simulation that when the distance between the center of masses reaches $\epsilon=0.1$, that distance decreases monotonically and quickly. Similarly, we must choose $t_c$ large enough to guarantee that the system has in fact reached a meta-stable state and to avoid a false turnaround, as illustrated in Fig \ref{fig:combineBAD}. Fig \ref{fig:combineTime} shows the time $t=t_c$ at which the two populations will be $\epsilon=0.1$ apart over varied parameter values based on direct simulations. In each panel, the curve defines a boundary. For example, if $\phi_0(0)$ is to the right of the curve in the first panel, then the two populations will be $\epsilon$ apart sooner than $t_c$. In each case, the curve becomes very steep near a critical parameter value. Consequently, as long we choose $t_c$ sufficiently large, our choice will not have much impact on the parameter value that defines our boundary. Guided by this reasoning, we choose to take as our condition for turnaround that the centers of the two populations are $\epsilon=0.1$ units away from one another at $t_c=500$.

To apply this condition, we solve a modification of system (\ref{eq:gauss2}) as a boundary value problem (BVP) with boundary condition $|\mu_b-\mu_\beta|=\epsilon=0.1$ at time $t=t_c=500.$ To satisfy all boundary conditions, we must consider one of the pertinent parameters as a stationary variable. For example, to determine the effect of $\chi_a$ on the transient behavior, we include the differential equation $\chi_a'=0$ in the BVP system. We then use the continuation software AUTO to solve this BVP across the selected parameter \cite{Bard}. The solution curve in parameter space defines a boundary between regions in which our model predicts that the bacteria turn around and in which it predicts that they combine. The results are shown in Fig \ref{fig:combine}.

\section*{Acknowledgments}
Hanna Salman acknowledges the support of the National Science Foundation (Grant PHY-1401576). Bard Ermentrout was supported by National Science Foundation Grant DMS--1219753.

%
%
%

\end{document}